\documentclass[%
 reprint,
 amsmath,amssymb,
 aps,
]{revtex4-2}

\usepackage{graphicx}% Include figure files
\usepackage{dcolumn}% Align table columns on decimal point
\usepackage{bm}% bold math
\usepackage{hyperref}% add hypertext capabilities
\usepackage{subcaption}
\usepackage{color}

\usepackage{caption}

\makeatletter
\newcommand*{\rom}[1]{\expandafter\@slowromancap\romannumeral #1@}
\makeatother
\begin{document}

\preprint{APS/123-QED}

\title{Thermal cycling - evidence for a generalized tunneling model and a tool to distinguish noise sources in quantum circuits}

\author{Yigal Reiss and Moshe Schechter}

\affiliation{Department of Physics, Ben-Gurion University of the Negev, Beer Sheva
84105, Israel}

\date{\today}

\begin{abstract}
Structural two level systems (TLSs) ubiquitous in amorphous solids are dramatically sensitive to thermal cycling to about $20$K and then back to low temperature, a process upon which the excitation energy of most TLSs is significantly changed. Using Monte Carlo simulations we demonstrate that this phenomenon is not contained within the standard tunneling model, but is well explained by a model that includes an additional set of TLSs that are pseudo-gapped at low energies, yet possess strong strain interaction through which they generate significant dynamical disorder upon thermal cycling. Our results provide additional support for the broad applicability of the Two-TLS model to amorphous solids at low temperatures, bringing us closer to a comprehensive understanding of the universal behavior of phonon attenuation in these materials. With regard to quantum superconducting circuits, our results suggest thermal cycling as a unique protocol to distinguish TLS noise from other noise sources. Possible relation of the Two-TLS model to ionizing radiation effects on long-time fluctuations in qubit relaxation times and on TLS scrambling is discussed.
\end{abstract}

\maketitle

Low temperature universality of phonon attenuation in amorphous solids has been a remarkable conundrum for over five decades \cite{zeller1971thermal,hunklinger1986thermal,pohl2002low}, as it signifies the role of the disordered state itself in dictating the low temperature characteristics of matter \cite{YuLeggett_1988}. While many of the corresponding experimental results can be explained as consequence of tunneling two-level systems (TLSs) as described by the standard tunneling model (STM) \cite{anderson1972anomalous,phillips1972tunneling,jackle1972ultrasonic}, some essential properties, including the microscopic nature of the TLSs and the origin of quantitative universality of phonon attenuation, cannot be understood within the STM.

Various generalizations of the STM were suggested over the years \cite{Karpov_1983, YuLeggett_1988, Galperin_1989, Buchenau_1992, parshin1994, solf1994two,Burin1995, Burin1996, Burin1998, lubchenko2001intrinsic, Kuhn_2003, Turlakov_2004, Parshin_2007, Vural_2011, Schechter_2013, Agarwal_2013, lubchenko2018, Carruzzo_2020}, addressing also the question of the low-temperature universality of phonon attenuation \cite{lubchenko2001intrinsic, Parshin_2007, Schechter_2013, Carruzzo_2020}.
Here we concentrate on one such suggestion, of a model consisting of two types of TLSs, distinct in the strength of their coupling to the strain \cite{Schechter_2013}. Within the Two-TLS model, in addition to the "standard" TLSs, which couple weakly to the strain, their exist a second type of TLSs having much stronger strain coupling. The ratio of the weak to strong strain coupling constants, $g \equiv \gamma_w/\gamma_s$ is the small dimensionless parameter of the model, its assumed value ${\approx}0.03$ given by the value of typical strain in strongly disordered systems \cite{Schechter_2008,Gaita_2011,Schechter_2013}.

In the past two decades, interest in the nature and characteristics of TLSs has broadened beyond the fundamental physics of amorphous solids, as TLSs were identified as a dominant source of noise and decoherence in quantum devices (\cite{Simmonds_2004,Shnirman_2005,Martinis_2005,Klimov_2018,Schlor_2019,Burnett_2019,Muller_2019}). Thus, the usefulness of the Two-TLS model, if applicable generically to amorphous solids, relates to the fundamental understanding of the low temperature characteristics of amorphous solids, and at the same time to better understanding of the source of noise, relaxation, and decoherence in quantum devices. With respect to the former, the smallness and universality of $g$ relates directly to the smallness and universality of phonon attenuation in amorphous solids, in that suggesting how the disordered state itself dictates universal phonon attenuation at low temperatures \cite{Schechter_2013}. In relation to the effects of TLSs on quantum devices, the Two-TLS model provides predictions which are distinct from those of the STM, some examples are given in Refs. \cite{Matityahu_2016,Schechter_2018,Yu_2022}, in that allowing better understanding of the deleterious effects of TLSs in quantum circuits, which may ultimately allow the design of protocols for their diminishment.

The Two-TLS model was developed in view of the similarities of the low temperature phenomena observed in amorphous solids and strongly disordered lattices. In the latter, TLSs can be classified according to their properties under inversion symmetry, providing the above distinction between weakly and strongly interacting TLSs, including the quantitative estimate of $g \approx 0.03$ \cite{Gaita_2011,Schechter_2013}. Indeed, the properties of the Two-TLS model, including the pseudo-gap of the strongly interacting TLSs at low energies, were verified in detail for KBr:CN, the archetypal disordered solid exhibiting the low temperature universality \cite{Gaita_2011,Churkin_2014}. Yet, recent results for the loss of superconducting resonators under fast bias sweep of the TLS energies give strong support for the applicability of the Two-TLS model to amorphous systems. In amorphous silicon, loss in excess of the maximal loss allowed by the standard TLSs as predicted by the STM is observed for fast bias sweep rates \cite{Yu_2022}. The bias sweeping creates a non-equilibrium state, where the excitation energies of strongly interacting TLSs are shifted to lower energies and contribute to the loss. Essential features of the experiment, including the dependence of the excess loss on the bias sweep rate, and the saturation of the excess loss only at very high resonator power, are in detailed agreement with the predictions of the Two-TLS model. Essentially the same results were recently obtained for AlOx resonators on Sapphire \cite{Y_Rosen_preparation}, suggesting the possible generality of the existence in amorphous solids of two types of TLSs distinct by their strain interaction strength.

A different experimental measurement that can distinguish between the STM and the Two-TLS model is that of the TLS energy spectrum as monitored by a superconducting qubit, and its change following thermal cycling \cite{PhysRevLett.105.177001}.
TLSs whose excitation energy is within the operation window of a qubit and their coupling to the qubit is larger than a certain threshold can be identified, and their excitation energy can be monitored over time. Experiments at cryogenic temperatures show that these TLS excitation energies fluctuate over time within a small energy interval, a result of their interaction with thermal TLSs. Over very long times, some TLSs experience a large shift in their excitation energy \cite{Klimov_2018,Schlor_2019}, yet, the TLS spectrum remains rather stable at low temperatures. Upon thermal cycling, i.e. heating the system to a certain temperature and then recooling to base temperature, many TLSs experience a small shift of their excitation energy, yet only few TLSs change their excitation energy considerably, say by more than $2$GHz ($\approx 0.1$K, the experimental window in Ref. \cite{PhysRevLett.105.177001})  when the heating is to ${\sim}1$K. However full rearrangement of the spectrum occurs when cycling the temperature to ${\sim}20$K \cite{PhysRevLett.105.177001} \cite{note_2024}. Within the STM such data are difficult to explain, as heating the system leads to the excitation of a large number of TLSs, but almost all of them return to their original state after cooling because their primary interaction is with the random field created by the static disorder, which remains unchanged at temperatures much smaller than the glass transition temperature. For the very few TLSs that do flip, their interaction with other TLSs is weak, resulting in a significant change in the excitation energies of only a negligible number of TLSs.
Within the Two-TLS model the situation is very different. Interactions of the strongly interacting ("$S$") TLSs are strong, and upon thermal cycling a process in which multiple TLSs (e.g. a strongly interacting S-TLS and a weakly interacting ("$\tau$") TLS) undergo excitation together may result in the survival of the flipped states upon re-cooling.
A flipped $S$-TLS then affects significantly the excitation energies of many $\tau$-TLSs. Within the Two-TLS model, the $S$-TLSs are pseudo-gapped, their energy dependent density of states (DOS) being small at low energies \cite{Schechter_2013,churkin2023strain}. The scale of ${\sim}20$K therefore relates to the energy where the DOS of the $S$-TLSs becomes appreciable, to affect upon thermal cycling most of the weakly interacting TLSs. At ${\sim}1$K only few $S$-TLSs are flipped upon thermal cycling, causing large energy shifts to only the few $\tau$-TLSs in their proximity, but still generating small shifts of the excitation energies of distant TLSs.

Here we quantify the above difference between the STM and the Two-TLS model by simulating the process of thermal cycling within both models. Measuring the number of TLSs that undergo a change in excitation energies larger than $0.1$K, we find that this number is negligible for the STM. Yet, for the Two-TLS model the number of $\tau$-TLSs, which correspond to the standard TLSs in the STM, that undergo a change in excitation energies larger than $0.1$K, changes
monotonously with temperature, from approximately $10\%$ for a heating temperature of $2$K to more than $90\%$ for a heating temperature of $20$K, following the change of the single particle DOS of the $S$-TLSs in equilibrium. Within the Two-TLS model we further find that a large fraction of the $\tau$-TLSs undergo smaller changes of their excitation energies upon thermal cycling already at temperatures ${\sim}1$K, in agreement with experiment \cite{PhysRevLett.105.177001}. Also this finding is not reproduced within the STM.
Thus we prove the relevance of Two-TLS model to the thermal cycling experiments in a striking contrast with a standard TLS model,
and in that we provide additional support for the generic applicability of the Two-TLS model to amorphous systems.

--- {\it The Two-TLS model}: The Two-TLS model assumes the existence of two types of TLSs, interacting weakly and strongly with strain, with the ratio of their strain interactions $g \equiv \gamma_w/\gamma_s$ being the dimensionless parameter of the model.
As a result of their strain interactions, one obtains a low-energy effective Hamiltonian for the TLSs \cite{Schechter_2013,churkin2023strain} with the form:

\begin{align}
\mathcal{H}_{s\tau} = -\sum_{i\neq j}[\frac{1}{2} J_{ij}^{ss}S_{i}S_{j} + J_{ij}^{s\tau}S_{i}\tau_{j} + \frac{1}{2} J_{ij}^{\tau \tau}\tau_{i}\tau_{j}]
\label{Eq:TwoTLS}
\end{align}
where
\begin{align}
J_{ij}^{ab} = c_{ij}^{ab} \cdot \frac{J_{0}^{ab}}{r_{ij}^{3}+\widetilde{a}^{3}} \, ,
\label{eq:interaction}
\end{align}
and a,b stand for $S$,$\tau$. The variables $S_{i} = \pm 1$ and $\tau_{i} = \pm 1$ represent the state of the TLSs. To avoid double summation, a factor of 1/2 is added. The coupling coefficient $J_{ij}$ depends on the distance $r_{ij}$ between the TLS, and $c_{ij}^{ab}$ is chosen randomly from a Gaussian distribution with a width of unity. To address the divergence at r=0, a cut-off parameter $\tilde{a}$ is introduced. The energy scales of the TLS-TLS interactions satisfy the relations
$J_{0}^{\tau\tau} = g \cdot J_{0}^{S\tau} = g^{2} \cdot J_{0}^{SS}$, where $J_{0}^{SS} \approx 300 K$ \cite{Schechter_2013}.
We note that the effective Hamiltonian may include also random field terms \cite{Schechter_2008,Schechter_2013}. However, these terms, not considered here, were shown \cite{Schechter_2013} to affect results only quantitatively and not appreciably, as they are of the same magnitude of the effective random field exerted by the high energy frozen S-TLSs \cite{Schechter_2008}. TLS tunneling is neglected in the Hamiltonian (\ref{Eq:TwoTLS}) because the tunneling amplitude is small compared to TLS – TLS interaction. TLS flips are modeled using Monte-Carlo (MC) tries with probability
given by the total energy of the initial and final states, which are well approximated by the Hamiltonian (\ref{Eq:TwoTLS}) (see Ref. \cite{churkin2023strain} for detailed considerations).

\begin{figure*}[t!]

\centering

\includegraphics[width=0.82\textwidth]{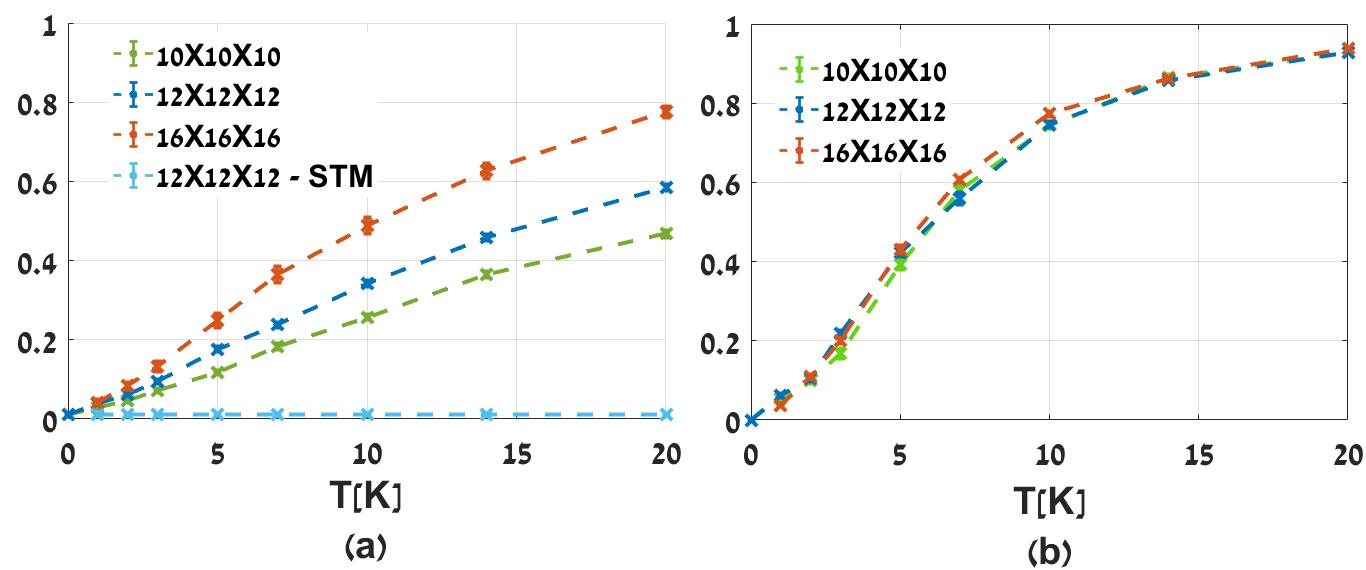}
\caption{Fraction of $\tau$-TLSs that change their excitation energy by more than $0.1$K upon thermal cycling, as function of the cycling temperature. (a) Direct calculation for the Two-TLS model with lattice sizes $L = 10, 12, 16$, and for the STM with $L=12$. In all simulations $\tilde{a} = 2$ and $\rho = 0.5$.
(b) Results for the Two-TLS model after extrapolations to large lattice sizes using the probability distribution $P_{s}$ generated from the simulations for $L_i = 10, 12, 16$. Results include only the number of $\tau$-TLSs that have a non-fluctuating change in their excitation energy (see Supplemental Material\cite{SupplementalMaterial} for details).
}
\label{Fig:0.1}
\end{figure*}

Within this model Hamiltonian, the single particle DOS of the S-TLSs
exhibits power law behavior for low energies, $\lesssim 10$K, where the exponent is dependent on the model's parameters, namely the cutoff $\tilde{a}$ and the spatial density \cite{churkin2023strain}. As a result, there is a notable disparity between the DOS of the $S$-TLSs at the energy ranges of $0-3$K and $10-20$K. A brief description of the Two-TLS model and its physical origin is given in the Supplemental Material\cite{SupplementalMaterial} (see also references \cite{Watson_1995,Liu_1998,Topp_1999,Topp_2002,POHL_opinion_1999,Treacy_2012} therein).

In contrast, the STM assumes only one type of TLS, where the approximate constant DOS at low energies is a result of a random field much larger than the TLS-TLS interactions. The effective Hamiltonian of this model takes the form:

\begin{align}
\mathcal{H}_{STM} = -\sum_{i} h_{i}\tau_{i} - \sum_{i\neq j}\frac{1}{2} J_{ij}^{\tau \tau}\tau_{i}\tau_{j} \, .
\label{Eq:interaction_random_field}
\end{align}
Here, $J_{ij}^{\tau \tau}$ is the same as in Eq. (\ref{eq:interaction}), and the magnitude of the random fields $h_{i}$ is discussed after Eq. (\ref{Eq:random}) below.

--- {\it Numerical simulation and results}: We perform MC simulations on cubic lattices of size $L^{3}$
within the model presented by the Hamiltonian in Eq. (\ref{Eq:TwoTLS}) with $\tilde{a} = 2$.
The TLSs are placed on lattice sites with a spatial density $\rho = 0.5$, and periodic boundary conditions are imposed.
Simulated annealing is conducted beginning with a random realization of spins at $300$K gradually reducing the temperature down to 0.02 K. At each of the $92$ temperatures $2000$ MC steps are executed to simulate the annealing process.
To optimize the low-energy state of the system, this process is being repeated $100$ times, starting from the realization at $40$K down to $0.02$K, and the state with lowest energy is selected as the initial state. The system is then heated using $10^{6}$ MC steps at the designated temperature, denoted henceforth the cycling temperature. Finally, the system is cooled down to $0.02$K with $2000$ MC steps for each temperature, following the same temperature ladder as the first cooling process.

To measure the difference between the initial realization and the final realization, we count the number of $\tau$-TLSs that differ in their excitation energy by more than $0.1$K, corresponding to the frequency window of the experiment \cite{PhysRevLett.105.177001}. Averaging is performed over $100$ such cycles per sample, for each of $N(L)$ samples, with $N(L)=250-2000$.

To simulate the STM we repeat the procedure above, only keeping the S-TLSs frozen at their annealed low energy state. This way the S-TLSs serve as a quenched random field acting on the $\tau$-TLSs, of magnitude

\begin{align}
h_{i} = \sum_{j}J_{ij}^{s\tau}\tilde{S}_{j}
\label{Eq:random}
\end{align}
where $\tilde{S}_{j}$ indicates the quenched configuration of the S-TLSs. The STM typically assumes that the magnitude of the random field is of order of the glass temperature \cite{anderson1972anomalous,phillips1972tunneling}. Our analysis assumes a much smaller random field, by a factor of $g$, in line with the predictions of the Two-TLS model \cite{Schechter_2008,Schechter_2013}. Thus, our results below give an upper limit to the changes in TLS energies upon thermal cycling within the STM.

Fig. \ref{Fig:0.1}(a) displays the fraction of $\tau$-TLSs whose excitation energy changes by more than $0.1$K upon thermal cycling, as a function of the cycling temperature. For the Two-TLS model, with dynamic S-TLSs, for system sizes $L=10,12,16$, we observe a monotonic increase in the said fraction of changed $\tau$-TLSs for all sample sizes. For the STM we observe very low number of changed TLSs for all temperatures, corresponding mostly to thermal TLSs which occasionally flip their spin, yet are not relevant to the experimental window (see also discussion in Supplemental Material\cite{SupplementalMaterial}).

We note that our protocol does not aim to reach equilibrium after temperature cycling, but rather an arbitrary low-energy metastable state, as it is exactly this behavior of the experimental system that results in the changes of TLS excitation energies upon thermal cycling. The two metastable states before and after thermal cycling are distinct, but statistically equivalent. Thus, while excitation energies of single $\tau$-TLSs change upon thermal cycling within the Two-TLS model, the averaged DOS of the $\tau$-TLSs as well as that of the $S$-TLSs remains unchanged after thermal cycling for both the STM and the Two-TLS model. To check the independence of our results on the specific cooling protocol and rate, we have repeated our calculations using $1000$ MC steps for each temperature, as well as with a protocol where the number of MC steps is enhanced with lowering the temperature up to $4000$ MC steps, all giving similar results.

\begin{figure}[t!]

\includegraphics[width=0.41\textwidth]{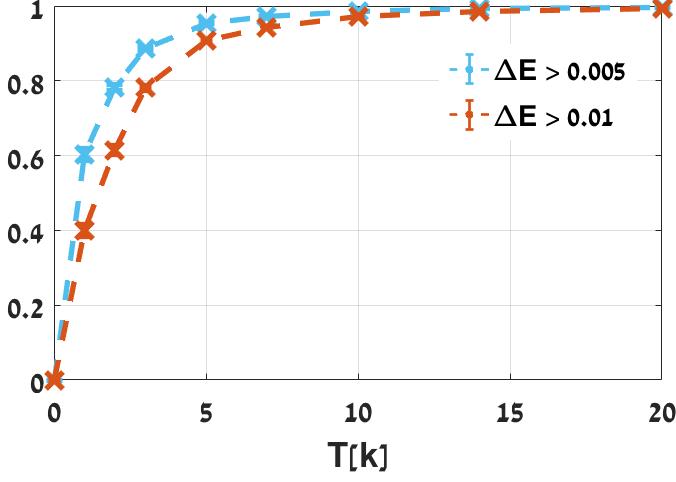}

\caption{Fraction of $\tau$-TLSs that have a non-fluctuating change in their excitation energy which exceeds $0.01$K and $0.005$K upon thermal cycling, as a function of the cycling temperature. The figure shows extrapolation results after convergence, using the probability distribution $P_{s}$ generated from the simulation at $L=12$. Within the STM, the number of TLSs that have a non-fluctuating change in their excitation energy which exceeds 0.005K is negligible.}

\label{Fig:0.01}
\end{figure}

Our results for the Two-TLS model clearly suffer from finite size effects. Indeed, the typical influence distance of an $S$-TLS that has flipped on the $\tau$-TLSs excitation energy for $\Delta E > 0.1$K is $R \approx 7$ (lattice spacings). In Fig. \ref{Fig:0.1}(b) we present our results for larger system sizes using the following extrapolation protocol:
First, we measure the number of S-TLSs that flip due to the heating and cooling process within our simulation of a given size $L_i$ and determine the discrete probability density function for the flip of $n$ TLSs in a single sample, $P_S(n)$.
 Then, we build a cube of length $L=k*L_i$ with periodic boundary conditions, where $L_i$ is the sample's length of the simulation from which $P_S (n)$ is deduced. The new sample, therefore, consists of $k^3$ cubes with a side length of $L_i$. For each small cube from the $k^3$ cubes, we first draw the number of flipped S-TLSs this cube contains from the discrete probability density function $P_S (n)$, and determine their position randomly in the cube. Finally, we measure the fraction of $\tau$-TLSs whose excitation energy changes by more than $\Delta E$ due to the flip of those $S$-TLSs.
Extrapolated results are presented for different initial sizes $L_i=10,12,16$, demonstrating the independence of the extrapolated results on the initial size $L_i$.
In the Supplemental Material\cite{SupplementalMaterial} we present further justification for the independence of the extrapolated results on sizes larger than $L_i = 10$, detailed results of the extrapolation for different sizes until convergence, and illustration of the limited sensitivity of the results to model parameters.

Repeating our measurement for the number of $\tau$-TLSs that experience, upon thermal cycling, changes in their excitation energies which are larger than a threshold smaller than $0.1$K, i.e. larger than $0.01$K and larger than $0.005$K, we find again that it monotonically increases with temperature. However, in agreement with experiment, the fraction of the TLS ensemble that experience such smaller excitation energy changes is appreciable already at low temperatures, of order of $1$K, see Fig. \ref{Fig:0.01}. This is since smaller changes in excitation energies of $\tau$-TLSs are a result of a flipped $S$-TLS at larger distances, and thus are more abundant at low temperatures, where the number of flipped $S$-TLSs upon thermal cycling is small.

{\it Discussion}: Whereas the generic presence of structural two-level systems in amorphous solids has been suggested over half a century ago, the characterization of their nature and the diminishment of their deleterious effects in low-temperature quantum devices has remained a formidable challenge. Phenomena such as long-time fluctuations in qubit relaxation times, detrimental to stability and scalability of quantum circuits, and loss of resonators under fast sweep bias rates which is in excess of that resulting from single type of TLSs as predicted by the STM, suggest the insufficiency of the STM to explain TLS nature and characteristics. A promising model towards bridging the gap in our understanding of TLSs and of the microscopic structure of amorphous solids is the Two-TLS model. This model suggests that in addition to the 'standard' TLSs as introduced by the STM, having a roughly homogenous single particle DOS and a very small ratio between their mutual interactions and the disorder field exerted on them, there exists a second type of TLSs. These '$S$'-TLSs have much stronger strain interaction and consequently TLS-TLS interactions of magnitude similar to the disorder field, yet are pseudo-gapped at low energies. They are subdominant in most phenomena in equilibrium at low temperatures, a result of their very small DOS at mK thermal energies and GHz resonant energies. Yet, upon switching their state, either over long times in equilibrium, or upon thermal cycling or non-equilibrium driving, they create significant dynamical disorder. Such a dynamical disorder results in non-equilibrium excess loss, as observed in Refs. \cite{Yu_2022,Y_Rosen_preparation} and in fluctuations of the TLS spectrum. These fluctuations over long times may result in the observed fluctuations in the relaxation times of qubits \cite{Klimov_2018,Schlor_2019}, and after thermal cycling they result in the re-initiation of the TLS spectrum, as observed experimentally in Ref. \cite{PhysRevLett.105.177001}, and shown theoretically here. Intriguingly, recent experimental data \cite{V_Iaia_preparation} find excess DOS of TLSs with a very large dipole moment that can be $S$-TLSs in a long-lived meta-stable state following driving of the system, which is yet partially reconfigured after thermal cycling to $1.5$K, and fully equilibrated after thermal cycling to $10$K, emphasizing the wide relevance of our results here.

Identifying the relevant noise sources in quantum circuits is a crucial step towards mitigating their effects. Here we suggest that thermal cycling can serve as a protocol to distinguish TLS noise from other source noises, as the change in TLS noise characteristics upon thermal cycling, with its characteristic functional form and typical energy scale of $\approx 10$K, is unique. Indeed, similar to our findings here, a strong change in noise characteristics of granular aluminum nanojunctions upon thermal cycling to $10$K \cite{Rieger2023}, and small changes of TLS excitation energies as measured by a superconducting resonator were observed upon thermal cycling to $0.3$K \cite{deGraaf2020}. Detailed studies of temperature dependence upon thermal cycling over a broad range could be useful in clarifying the source of noise in these and other systems.

A major obstacle towards stability and scalability of quantum superconducting circuits is the long-time fluctuations in qubit relaxation times \cite{Klimov_2018,Schlor_2019}. Recently it was argued that these rare fluctuations are initiated by ionizing radiation \cite{Grunhaupt2018,Vepsalainen2020,Wilen2021,Thorbeck2023}, and that ionizing radiation events are related to scrambling of the TLS bath \cite{Thorbeck2023}. An ionizing radiation event inputs a large amount of energy into the system at a given time, which is then dissipated, resembling thermal cycling protocol only over a very short time. The question still remains of how this process produces long time fluctuations in qubit properties, and its relation to the scrambling of TLSs. As our work here shows that the presence of strongly interacting TLSs results in dynamical disorder upon raising and lowering the energy of the system, it would be of much interest to study the relation of the Two-TLS model to the scrambling of the TLS spectrum and to the long time fluctuations in qubit characteristics following ionizing radiation events.

The experimental and theoretical verification of the generic existence of two types of TLSs in amorphous solids, differing by the magnitude of their interaction with the strain, and possibly by the magnitude of their dipole moment, would be a significant advance of our understanding of the nature of TLSs in amorphous solids; and with that the characteristics of noise, relaxation, decoherence, and loss in quantum circuits, and the essential physics of the low temperature universality in amorphous solids.
Detection of the two types of TLSs through the dielectric loss of resonators under electric bias sweep \cite{Yu_2022} requires the strongly interacting TLSs to have also a large electric dipole moment. Detection of strongly interacting TLSs not having large electric dipoles in similar bias sweep experiments requires coupling to their elastic moment, by measuring internal friction and using an acoustic bias sweep field. Here we suggest an alternative for the detection of strongly interacting TLSs in generic amorphous solids, through thermal cycling experiments. While initialization of the TLS bath by thermal cycling is a routine procedure, careful data collecting of TLS statistics as reported by Shalibo et. al. \cite{PhysRevLett.105.177001}, performed over variety of materials, and compared against theory, can further probe the general existence of two types of TLSs in amorphous solids, differing essentially by their interaction with the strain, as suggested by the Two-TLS model. Furthermore, taking advantage of the quantitative dependence of our results on the parameters of the Two-TLS model (see details in Supplemental Material\cite{SupplementalMaterial}), comprehensive study of the statistics of the changes of TLS excitation energies following thermal cycling to different temperatures would allow the extraction of the parameters of the Two-TLS model.

We would like to thank Alexander Burin for useful discussions. M.S. acknowledges support by the Israel Science Foundation (Grant No. 2300/19 and Grant No. 3679/24).

\bibliographystyle{apsrev4-2-titles}
\bibliography{reff_PRB_resubmission}

\providecommand{\noopsort}[1]{}\providecommand{\singleletter}[1]{#1}%
\begin{thebibliography}{55}%
\makeatletter
\providecommand \@ifxundefined [1]{%
 \@ifx{#1\undefined}
}%
\providecommand \@ifnum [1]{%
 \ifnum #1\expandafter \@firstoftwo
 \else \expandafter \@secondoftwo
 \fi
}%
\providecommand \@ifx [1]{%
 \ifx #1\expandafter \@firstoftwo
 \else \expandafter \@secondoftwo
 \fi
}%
\providecommand \natexlab [1]{#1}%
\providecommand \enquote  [1]{``#1''}%
\providecommand \bibnamefont  [1]{#1}%
\providecommand \bibfnamefont [1]{#1}%
\providecommand \citenamefont [1]{#1}%
\providecommand \href@noop [0]{\@secondoftwo}%
\providecommand \href [0]{\begingroup \@sanitize@url \@href}%
\providecommand \@href[1]{\@@startlink{#1}\@@href}%
\providecommand \@@href[1]{\endgroup#1\@@endlink}%
\providecommand \@sanitize@url [0]{\catcode `\\12\catcode `\$12\catcode
  `\&12\catcode `\#12\catcode `\^12\catcode `\_12\catcode `\%12\relax}%
\providecommand \@@startlink[1]{}%
\providecommand \@@endlink[0]{}%
\providecommand \url  [0]{\begingroup\@sanitize@url \@url }%
\providecommand \@url [1]{\endgroup\@href {#1}{\urlprefix }}%
\providecommand \urlprefix  [0]{URL }%
\providecommand \Eprint [0]{\href }%
\providecommand \doibase [0]{https://doi.org/}%
\providecommand \selectlanguage [0]{\@gobble}%
\providecommand \bibinfo  [0]{\@secondoftwo}%
\providecommand \bibfield  [0]{\@secondoftwo}%
\providecommand \translation [1]{[#1]}%
\providecommand \BibitemOpen [0]{}%
\providecommand \bibitemStop [0]{}%
\providecommand \bibitemNoStop [0]{.\EOS\space}%
\providecommand \EOS [0]{\spacefactor3000\relax}%
\providecommand \BibitemShut  [1]{\csname bibitem#1\endcsname}%
\let\auto@bib@innerbib\@empty
%</preamble>
\bibitem [{\citenamefont {Zeller}\ and\ \citenamefont
  {Pohl}(1971)}]{zeller1971thermal}%
  \BibitemOpen
  \bibfield  {author} {\bibinfo {author} {\bibfnamefont {R.}~\bibnamefont
  {Zeller}}\ and\ \bibinfo {author} {\bibfnamefont {R.}~\bibnamefont {Pohl}},\
  }\bibfield  {title} {\emph {\bibinfo {title} {Thermal conductivity and
  specific heat of noncrystalline solids}},\ }\href@noop {} {\bibfield
  {journal} {\bibinfo  {journal} {Physical Review B}\ }\textbf {\bibinfo
  {volume} {4}},\ \bibinfo {pages} {2029} (\bibinfo {year} {1971})}\BibitemShut
  {NoStop}%
\bibitem [{\citenamefont {Hunklinger}\ and\ \citenamefont
  {Raychaudhuri}(1986)}]{hunklinger1986thermal}%
  \BibitemOpen
  \bibfield  {author} {\bibinfo {author} {\bibfnamefont {S.}~\bibnamefont
  {Hunklinger}}\ and\ \bibinfo {author} {\bibfnamefont {A.~K.}\ \bibnamefont
  {Raychaudhuri}},\ }\bibinfo {title} {Thermal and elastic anomalies in glasses
  at low temperatures},\ in\ \href@noop {} {\emph {\bibinfo {booktitle}
  {Progress in Low Temperature Physics}}},\ Vol.~\bibinfo {volume} {9}\
  (\bibinfo  {publisher} {Elsevier},\ \bibinfo {year} {1986})\ pp.\ \bibinfo
  {pages} {265--344}\BibitemShut {NoStop}%
\bibitem [{\citenamefont {Pohl}\ \emph {et~al.}(2002)\citenamefont {Pohl},
  \citenamefont {Liu},\ and\ \citenamefont {Thompson}}]{pohl2002low}%
  \BibitemOpen
  \bibfield  {author} {\bibinfo {author} {\bibfnamefont {R.~O.}\ \bibnamefont
  {Pohl}}, \bibinfo {author} {\bibfnamefont {X.}~\bibnamefont {Liu}},\ and\
  \bibinfo {author} {\bibfnamefont {E.}~\bibnamefont {Thompson}},\ }\bibfield
  {title} {\emph {\bibinfo {title} {Low-temperature thermal conductivity and
  acoustic attenuation in amorphous solids}},\ }\href@noop {} {\bibfield
  {journal} {\bibinfo  {journal} {Reviews of Modern Physics}\ }\textbf
  {\bibinfo {volume} {74}},\ \bibinfo {pages} {991} (\bibinfo {year}
  {2002})}\BibitemShut {NoStop}%
\bibitem [{\citenamefont {Yu}\ and\ \citenamefont
  {Leggett}(1988)}]{YuLeggett_1988}%
  \BibitemOpen
  \bibfield  {author} {\bibinfo {author} {\bibfnamefont {C.~C.}\ \bibnamefont
  {Yu}}\ and\ \bibinfo {author} {\bibfnamefont {A.~J.}\ \bibnamefont
  {Leggett}},\ }\bibfield  {title} {\emph {\bibinfo {title} {Low temperature
  properties of amorphous materials: Through a glass darkly}},\ }\href
  {https://doi.org/10.1063/1.341527} {\bibfield  {journal} {\bibinfo  {journal}
  {Comments on Condensed Matter Physics}\ }\textbf {\bibinfo {volume} {14}},\
  \bibinfo {pages} {231--251} (\bibinfo {year} {1988})}\BibitemShut {NoStop}%
\bibitem [{\citenamefont {Anderson}\ \emph {et~al.}(1972)\citenamefont
  {Anderson}, \citenamefont {Halperin},\ and\ \citenamefont
  {Varma}}]{anderson1972anomalous}%
  \BibitemOpen
  \bibfield  {author} {\bibinfo {author} {\bibfnamefont {P.~W.}\ \bibnamefont
  {Anderson}}, \bibinfo {author} {\bibfnamefont {B.~I.}\ \bibnamefont
  {Halperin}},\ and\ \bibinfo {author} {\bibfnamefont {C.~M.}\ \bibnamefont
  {Varma}},\ }\bibfield  {title} {\emph {\bibinfo {title} {Anomalous
  low-temperature thermal properties of glasses and spin glasses}},\
  }\href@noop {} {\bibfield  {journal} {\bibinfo  {journal} {Philosophical
  Magazine}\ }\textbf {\bibinfo {volume} {25}},\ \bibinfo {pages} {1--9}
  (\bibinfo {year} {1972})}\BibitemShut {NoStop}%
\bibitem [{\citenamefont {Phillips}(1972)}]{phillips1972tunneling}%
  \BibitemOpen
  \bibfield  {author} {\bibinfo {author} {\bibfnamefont {W.~A.}\ \bibnamefont
  {Phillips}},\ }\bibfield  {title} {\emph {\bibinfo {title} {Tunneling states
  in amorphous solids}},\ }\href@noop {} {\bibfield  {journal} {\bibinfo
  {journal} {Journal of low temperature physics}\ }\textbf {\bibinfo {volume}
  {7}},\ \bibinfo {pages} {351--360} (\bibinfo {year} {1972})}\BibitemShut
  {NoStop}%
\bibitem [{\citenamefont {J{\"a}ckle}(1972)}]{jackle1972ultrasonic}%
  \BibitemOpen
  \bibfield  {author} {\bibinfo {author} {\bibfnamefont {J.}~\bibnamefont
  {J{\"a}ckle}},\ }\bibfield  {title} {\emph {\bibinfo {title} {On the
  ultrasonic attenuation in glasses at low temperatures}},\ }\href@noop {}
  {\bibfield  {journal} {\bibinfo  {journal} {Zeitschrift f{\"u}r Physik A
  Hadrons and nuclei}\ }\textbf {\bibinfo {volume} {257}},\ \bibinfo {pages}
  {212--223} (\bibinfo {year} {1972})}\BibitemShut {NoStop}%
\bibitem [{\citenamefont {Karpov}\ \emph {et~al.}(1983)\citenamefont {Karpov},
  \citenamefont {Klinger},\ and\ \citenamefont {Ignat'ev}}]{Karpov_1983}%
  \BibitemOpen
  \bibfield  {author} {\bibinfo {author} {\bibfnamefont {V.~G.}\ \bibnamefont
  {Karpov}}, \bibinfo {author} {\bibfnamefont {M.~I.}\ \bibnamefont
  {Klinger}},\ and\ \bibinfo {author} {\bibfnamefont {F.~N.}\ \bibnamefont
  {Ignat'ev}},\ }\bibfield  {title} {\emph {\bibinfo {title} {Theory of the
  low-temperature anomalies in the thermal properties of amorphous
  structures}},\ }\href@noop {} {\bibfield  {journal} {\bibinfo  {journal}
  {Soviet Physics JETP}\ }\textbf {\bibinfo {volume} {57}},\ \bibinfo {pages}
  {439--448} (\bibinfo {year} {1983})}\BibitemShut {NoStop}%
\bibitem [{\citenamefont {Yu.M.~Galperin}\ and\ \citenamefont
  {Kozub}(1989)}]{Galperin_1989}%
  \BibitemOpen
  \bibfield  {author} {\bibinfo {author} {\bibfnamefont {V.~K.}\ \bibnamefont
  {Yu.M.~Galperin}}\ and\ \bibinfo {author} {\bibfnamefont {V.}~\bibnamefont
  {Kozub}},\ }\bibfield  {title} {\emph {\bibinfo {title} {Localized states in
  glasses}},\ }\href {https://doi.org/10.1080/00018738900101162} {\bibfield
  {journal} {\bibinfo  {journal} {Advances in Physics}\ }\textbf {\bibinfo
  {volume} {38}},\ \bibinfo {pages} {669--737} (\bibinfo {year}
  {1989})}\BibitemShut {NoStop}%
\bibitem [{\citenamefont {Buchenau}\ \emph {et~al.}(1992)\citenamefont
  {Buchenau}, \citenamefont {Galperin}, \citenamefont {Gurevich}, \citenamefont
  {Parshin}, \citenamefont {Ramos},\ and\ \citenamefont
  {Schober}}]{Buchenau_1992}%
  \BibitemOpen
  \bibfield  {author} {\bibinfo {author} {\bibfnamefont {U.}~\bibnamefont
  {Buchenau}}, \bibinfo {author} {\bibfnamefont {Y.~M.}\ \bibnamefont
  {Galperin}}, \bibinfo {author} {\bibfnamefont {V.~L.}\ \bibnamefont
  {Gurevich}}, \bibinfo {author} {\bibfnamefont {D.~A.}\ \bibnamefont
  {Parshin}}, \bibinfo {author} {\bibfnamefont {M.~A.}\ \bibnamefont {Ramos}},\
  and\ \bibinfo {author} {\bibfnamefont {H.~R.}\ \bibnamefont {Schober}},\
  }\bibfield  {title} {\emph {\bibinfo {title} {Interaction of soft modes and
  sound waves in glasses}},\ }\href {https://doi.org/10.1103/PhysRevB.46.2798}
  {\bibfield  {journal} {\bibinfo  {journal} {Phys. Rev. B}\ }\textbf {\bibinfo
  {volume} {46}},\ \bibinfo {pages} {2798--2808} (\bibinfo {year}
  {1992})}\BibitemShut {NoStop}%
\bibitem [{\citenamefont {Parshin}(1994)}]{parshin1994}%
  \BibitemOpen
  \bibfield  {author} {\bibinfo {author} {\bibfnamefont {D.}~\bibnamefont
  {Parshin}},\ }\bibfield  {title} {\emph {\bibinfo {title} {Interactions of
  soft atomic potentials and universality of low-temperature properties of
  glasses}},\ }\href@noop {} {\bibfield  {journal} {\bibinfo  {journal}
  {Physical Review B}\ }\textbf {\bibinfo {volume} {49}},\ \bibinfo {pages}
  {9400} (\bibinfo {year} {1994})}\BibitemShut {NoStop}%
\bibitem [{\citenamefont {Solf}\ and\ \citenamefont
  {Klein}(1994)}]{solf1994two}%
  \BibitemOpen
  \bibfield  {author} {\bibinfo {author} {\bibfnamefont {M.~P.}\ \bibnamefont
  {Solf}}\ and\ \bibinfo {author} {\bibfnamefont {M.~W.}\ \bibnamefont
  {Klein}},\ }\bibfield  {title} {\emph {\bibinfo {title} {Two-level tunneling
  states and the constant density of states in quadrupolar glasses}},\
  }\href@noop {} {\bibfield  {journal} {\bibinfo  {journal} {Physical Review
  B}\ }\textbf {\bibinfo {volume} {49}},\ \bibinfo {pages} {12703} (\bibinfo
  {year} {1994})}\BibitemShut {NoStop}%
\bibitem [{\citenamefont {Burin}(1995)}]{Burin1995}%
  \BibitemOpen
  \bibfield  {author} {\bibinfo {author} {\bibfnamefont {A.~L.}\ \bibnamefont
  {Burin}},\ }\bibfield  {title} {\emph {\bibinfo {title} {Dipole gap effects
  in low energy excitation spectrum of amorphous solids. theory for dielectric
  relaxation}},\ }\href {https://doi.org/10.1007/BF00751512} {\bibfield
  {journal} {\bibinfo  {journal} {Journal of Low Temperature Physics}\ }\textbf
  {\bibinfo {volume} {100}},\ \bibinfo {pages} {309--337} (\bibinfo {year}
  {1995})}\BibitemShut {NoStop}%
\bibitem [{\citenamefont {Burin}\ and\ \citenamefont
  {Kagan}(1996)}]{Burin1996}%
  \BibitemOpen
  \bibfield  {author} {\bibinfo {author} {\bibfnamefont {A.~L.}\ \bibnamefont
  {Burin}}\ and\ \bibinfo {author} {\bibfnamefont {Y.}~\bibnamefont {Kagan}},\
  }\bibfield  {title} {\emph {\bibinfo {title} {On the nature of the universal
  properties of amorphous solids}},\ }\href
  {https://doi.org/10.1007/BF02571128} {\bibfield  {journal} {\bibinfo
  {journal} {Czechoslovak Journal of Physics}\ }\textbf {\bibinfo {volume}
  {46}},\ \bibinfo {pages} {2273--2274} (\bibinfo {year} {1996})}\BibitemShut
  {NoStop}%
\bibitem [{\citenamefont {Burin}\ \emph {et~al.}(1998)\citenamefont {Burin},
  \citenamefont {Natelson}, \citenamefont {Osheroff},\ and\ \citenamefont
  {Kagan}}]{Burin1998}%
  \BibitemOpen
  \bibfield  {author} {\bibinfo {author} {\bibfnamefont {A.~L.}\ \bibnamefont
  {Burin}}, \bibinfo {author} {\bibfnamefont {D.}~\bibnamefont {Natelson}},
  \bibinfo {author} {\bibfnamefont {D.~D.}\ \bibnamefont {Osheroff}},\ and\
  \bibinfo {author} {\bibfnamefont {Y.}~\bibnamefont {Kagan}},\ }\bibinfo
  {title} {Interactions between tunneling defects in amorphous solids},\ in\
  \href {https://doi.org/10.1007/978-3-662-03695-2_5} {\emph {\bibinfo
  {booktitle} {Tunneling Systems in Amorphous and Crystalline Solids}}},\
  \bibinfo {editor} {edited by\ \bibinfo {editor} {\bibfnamefont
  {P.}~\bibnamefont {Esquinazi}}}\ (\bibinfo  {publisher} {Springer Berlin
  Heidelberg},\ \bibinfo {address} {Berlin, Heidelberg},\ \bibinfo {year}
  {1998})\ pp.\ \bibinfo {pages} {223--315}\BibitemShut {NoStop}%
\bibitem [{\citenamefont {Lubchenko}\ and\ \citenamefont
  {Wolynes}(2001)}]{lubchenko2001intrinsic}%
  \BibitemOpen
  \bibfield  {author} {\bibinfo {author} {\bibfnamefont {V.}~\bibnamefont
  {Lubchenko}}\ and\ \bibinfo {author} {\bibfnamefont {P.~G.}\ \bibnamefont
  {Wolynes}},\ }\bibfield  {title} {\emph {\bibinfo {title} {Intrinsic quantum
  excitations of low temperature glasses}},\ }\href@noop {} {\bibfield
  {journal} {\bibinfo  {journal} {Physical Review Letters}\ }\textbf {\bibinfo
  {volume} {87}},\ \bibinfo {pages} {195901} (\bibinfo {year}
  {2001})}\BibitemShut {NoStop}%
\bibitem [{\citenamefont {Kuhn}(2003)}]{Kuhn_2003}%
  \BibitemOpen
  \bibfield  {author} {\bibinfo {author} {\bibfnamefont {R.}~\bibnamefont
  {Kuhn}},\ }\bibfield  {title} {\emph {\bibinfo {title} {Universality in
  glassy low-temperature physics}},\ }\href
  {https://doi.org/10.1209/epl/i2003-00397-8} {\bibfield  {journal} {\bibinfo
  {journal} {Europhysics Letters}\ }\textbf {\bibinfo {volume} {62}},\ \bibinfo
  {pages} {313} (\bibinfo {year} {2003})}\BibitemShut {NoStop}%
\bibitem [{\citenamefont {Turlakov}(2004)}]{Turlakov_2004}%
  \BibitemOpen
  \bibfield  {author} {\bibinfo {author} {\bibfnamefont {M.}~\bibnamefont
  {Turlakov}},\ }\bibfield  {title} {\emph {\bibinfo {title} {Universal sound
  absorption in low-temperature amorphous solids}},\ }\href
  {https://doi.org/10.1103/PhysRevLett.93.035501} {\bibfield  {journal}
  {\bibinfo  {journal} {Phys. Rev. Lett.}\ }\textbf {\bibinfo {volume} {93}},\
  \bibinfo {pages} {035501} (\bibinfo {year} {2004})}\BibitemShut {NoStop}%
\bibitem [{\citenamefont {Parshin}\ \emph {et~al.}(2007)\citenamefont
  {Parshin}, \citenamefont {Schober},\ and\ \citenamefont
  {Gurevich}}]{Parshin_2007}%
  \BibitemOpen
  \bibfield  {author} {\bibinfo {author} {\bibfnamefont {D.~A.}\ \bibnamefont
  {Parshin}}, \bibinfo {author} {\bibfnamefont {H.~R.}\ \bibnamefont
  {Schober}},\ and\ \bibinfo {author} {\bibfnamefont {V.~L.}\ \bibnamefont
  {Gurevich}},\ }\bibfield  {title} {\emph {\bibinfo {title} {Vibrational
  instability, two-level systems, and the boson peak in glasses}},\ }\href
  {https://doi.org/10.1103/PhysRevB.76.064206} {\bibfield  {journal} {\bibinfo
  {journal} {Phys. Rev. B}\ }\textbf {\bibinfo {volume} {76}},\ \bibinfo
  {pages} {064206} (\bibinfo {year} {2007})}\BibitemShut {NoStop}%
\bibitem [{\citenamefont {Vural}\ and\ \citenamefont
  {Leggett}(2011)}]{Vural_2011}%
  \BibitemOpen
  \bibfield  {author} {\bibinfo {author} {\bibfnamefont {D.~C.}\ \bibnamefont
  {Vural}}\ and\ \bibinfo {author} {\bibfnamefont {A.~J.}\ \bibnamefont
  {Leggett}},\ }\bibfield  {title} {\emph {\bibinfo {title} {Universal sound
  absorption in amorphous solids: A theory of elastically coupled generic
  blocks}},\ }\href
  {https://doi.org/https://doi.org/10.1016/j.jnoncrysol.2011.06.035} {\bibfield
   {journal} {\bibinfo  {journal} {Journal of Non-Crystalline Solids}\ }\textbf
  {\bibinfo {volume} {357}},\ \bibinfo {pages} {3528--3537} (\bibinfo {year}
  {2011})}\BibitemShut {NoStop}%
\bibitem [{\citenamefont {Schechter}\ and\ \citenamefont
  {Stamp}(2013)}]{Schechter_2013}%
  \BibitemOpen
  \bibfield  {author} {\bibinfo {author} {\bibfnamefont {M.}~\bibnamefont
  {Schechter}}\ and\ \bibinfo {author} {\bibfnamefont {P.~C.~E.}\ \bibnamefont
  {Stamp}},\ }\bibfield  {title} {\emph {\bibinfo {title} {Inversion symmetric
  two-level systems and the low-temperature universality in disordered
  solids}},\ }\href {https://doi.org/10.1103/PhysRevB.88.174202} {\bibfield
  {journal} {\bibinfo  {journal} {Phys. Rev. B}\ }\textbf {\bibinfo {volume}
  {88}},\ \bibinfo {pages} {174202} (\bibinfo {year} {2013})}\BibitemShut
  {NoStop}%
\bibitem [{\citenamefont {Agarwal}\ \emph {et~al.}(2013)\citenamefont
  {Agarwal}, \citenamefont {Martin}, \citenamefont {Lukin},\ and\ \citenamefont
  {Demler}}]{Agarwal_2013}%
  \BibitemOpen
  \bibfield  {author} {\bibinfo {author} {\bibfnamefont {K.}~\bibnamefont
  {Agarwal}}, \bibinfo {author} {\bibfnamefont {I.}~\bibnamefont {Martin}},
  \bibinfo {author} {\bibfnamefont {M.~D.}\ \bibnamefont {Lukin}},\ and\
  \bibinfo {author} {\bibfnamefont {E.}~\bibnamefont {Demler}},\ }\bibfield
  {title} {\emph {\bibinfo {title} {Polaronic model of two-level systems in
  amorphous solids}},\ }\href {https://doi.org/10.1103/PhysRevB.87.144201}
  {\bibfield  {journal} {\bibinfo  {journal} {Phys. Rev. B}\ }\textbf {\bibinfo
  {volume} {87}},\ \bibinfo {pages} {144201} (\bibinfo {year}
  {2013})}\BibitemShut {NoStop}%
\bibitem [{\citenamefont {Lubchenko}(2018)}]{lubchenko2018}%
  \BibitemOpen
  \bibfield  {author} {\bibinfo {author} {\bibfnamefont {V.}~\bibnamefont
  {Lubchenko}},\ }\bibfield  {title} {\emph {\bibinfo {title} {Low-temperature
  anomalies in disordered solids: a cold case of contested relics?}},\
  }\href@noop {} {\bibfield  {journal} {\bibinfo  {journal} {Advances in
  Physics: X}\ }\textbf {\bibinfo {volume} {3}},\ \bibinfo {pages} {1510296}
  (\bibinfo {year} {2018})}\BibitemShut {NoStop}%
\bibitem [{\citenamefont {Carruzzo}\ and\ \citenamefont
  {Yu}(2020)}]{Carruzzo_2020}%
  \BibitemOpen
  \bibfield  {author} {\bibinfo {author} {\bibfnamefont {H.~M.}\ \bibnamefont
  {Carruzzo}}\ and\ \bibinfo {author} {\bibfnamefont {C.~C.}\ \bibnamefont
  {Yu}},\ }\bibfield  {title} {\emph {\bibinfo {title} {Why phonon scattering
  in glasses is universally small at low temperatures}},\ }\href
  {https://doi.org/10.1103/PhysRevLett.124.075902} {\bibfield  {journal}
  {\bibinfo  {journal} {Phys. Rev. Lett.}\ }\textbf {\bibinfo {volume} {124}},\
  \bibinfo {pages} {075902} (\bibinfo {year} {2020})}\BibitemShut {NoStop}%
\bibitem [{\citenamefont {Schechter}\ and\ \citenamefont
  {Stamp}(2008)}]{Schechter_2008}%
  \BibitemOpen
  \bibfield  {author} {\bibinfo {author} {\bibfnamefont {M.}~\bibnamefont
  {Schechter}}\ and\ \bibinfo {author} {\bibfnamefont {P.~C.~E.}\ \bibnamefont
  {Stamp}},\ }\bibfield  {title} {\emph {\bibinfo {title} {What are the
  interactions in quantum glasses?}},\ }\href
  {https://doi.org/10.1088/0953-8984/20/24/244136} {\bibfield  {journal}
  {\bibinfo  {journal} {Journal of Physics: Condensed Matter}\ }\textbf
  {\bibinfo {volume} {20}},\ \bibinfo {pages} {244136} (\bibinfo {year}
  {2008})}\BibitemShut {NoStop}%
\bibitem [{\citenamefont {Gaita-Ari\~no}\ and\ \citenamefont
  {Schechter}(2011)}]{Gaita_2011}%
  \BibitemOpen
  \bibfield  {author} {\bibinfo {author} {\bibfnamefont {A.}~\bibnamefont
  {Gaita-Ari\~no}}\ and\ \bibinfo {author} {\bibfnamefont {M.}~\bibnamefont
  {Schechter}},\ }\bibfield  {title} {\emph {\bibinfo {title} {Identification
  of strong and weak interacting two-level systems in kbr:cn}},\ }\href
  {https://doi.org/10.1103/PhysRevLett.107.105504} {\bibfield  {journal}
  {\bibinfo  {journal} {Phys. Rev. Lett.}\ }\textbf {\bibinfo {volume} {107}},\
  \bibinfo {pages} {105504} (\bibinfo {year} {2011})}\BibitemShut {NoStop}%
\bibitem [{\citenamefont {Simmonds}\ \emph {et~al.}(2004)\citenamefont
  {Simmonds}, \citenamefont {Lang}, \citenamefont {Hite}, \citenamefont {Nam},
  \citenamefont {Pappas},\ and\ \citenamefont {Martinis}}]{Simmonds_2004}%
  \BibitemOpen
  \bibfield  {author} {\bibinfo {author} {\bibfnamefont {R.~W.}\ \bibnamefont
  {Simmonds}}, \bibinfo {author} {\bibfnamefont {K.~M.}\ \bibnamefont {Lang}},
  \bibinfo {author} {\bibfnamefont {D.~A.}\ \bibnamefont {Hite}}, \bibinfo
  {author} {\bibfnamefont {S.}~\bibnamefont {Nam}}, \bibinfo {author}
  {\bibfnamefont {D.~P.}\ \bibnamefont {Pappas}},\ and\ \bibinfo {author}
  {\bibfnamefont {J.~M.}\ \bibnamefont {Martinis}},\ }\bibfield  {title} {\emph
  {\bibinfo {title} {Decoherence in josephson phase qubits from junction
  resonators}},\ }\href {https://doi.org/10.1103/PhysRevLett.93.077003}
  {\bibfield  {journal} {\bibinfo  {journal} {Phys. Rev. Lett.}\ }\textbf
  {\bibinfo {volume} {93}},\ \bibinfo {pages} {077003} (\bibinfo {year}
  {2004})}\BibitemShut {NoStop}%
\bibitem [{\citenamefont {Shnirman}\ \emph {et~al.}(2005)\citenamefont
  {Shnirman}, \citenamefont {Sch\"on}, \citenamefont {Martin},\ and\
  \citenamefont {Makhlin}}]{Shnirman_2005}%
  \BibitemOpen
  \bibfield  {author} {\bibinfo {author} {\bibfnamefont {A.}~\bibnamefont
  {Shnirman}}, \bibinfo {author} {\bibfnamefont {G.}~\bibnamefont {Sch\"on}},
  \bibinfo {author} {\bibfnamefont {I.}~\bibnamefont {Martin}},\ and\ \bibinfo
  {author} {\bibfnamefont {Y.}~\bibnamefont {Makhlin}},\ }\bibfield  {title}
  {\emph {\bibinfo {title} {Low- and high-frequency noise from coherent
  two-level systems}},\ }\href {https://doi.org/10.1103/PhysRevLett.94.127002}
  {\bibfield  {journal} {\bibinfo  {journal} {Phys. Rev. Lett.}\ }\textbf
  {\bibinfo {volume} {94}},\ \bibinfo {pages} {127002} (\bibinfo {year}
  {2005})}\BibitemShut {NoStop}%
\bibitem [{\citenamefont {Martinis}\ \emph {et~al.}(2005)\citenamefont
  {Martinis}, \citenamefont {Cooper}, \citenamefont {McDermott}, \citenamefont
  {Steffen}, \citenamefont {Ansmann}, \citenamefont {Osborn}, \citenamefont
  {Cicak}, \citenamefont {Oh}, \citenamefont {Pappas}, \citenamefont
  {Simmonds},\ and\ \citenamefont {Yu}}]{Martinis_2005}%
  \BibitemOpen
  \bibfield  {author} {\bibinfo {author} {\bibfnamefont {J.~M.}\ \bibnamefont
  {Martinis}}, \bibinfo {author} {\bibfnamefont {K.~B.}\ \bibnamefont
  {Cooper}}, \bibinfo {author} {\bibfnamefont {R.}~\bibnamefont {McDermott}},
  \bibinfo {author} {\bibfnamefont {M.}~\bibnamefont {Steffen}}, \bibinfo
  {author} {\bibfnamefont {M.}~\bibnamefont {Ansmann}}, \bibinfo {author}
  {\bibfnamefont {K.~D.}\ \bibnamefont {Osborn}}, \bibinfo {author}
  {\bibfnamefont {K.}~\bibnamefont {Cicak}}, \bibinfo {author} {\bibfnamefont
  {S.}~\bibnamefont {Oh}}, \bibinfo {author} {\bibfnamefont {D.~P.}\
  \bibnamefont {Pappas}}, \bibinfo {author} {\bibfnamefont {R.~W.}\
  \bibnamefont {Simmonds}},\ and\ \bibinfo {author} {\bibfnamefont {C.~C.}\
  \bibnamefont {Yu}},\ }\bibfield  {title} {\emph {\bibinfo {title}
  {Decoherence in josephson qubits from dielectric loss}},\ }\href
  {https://doi.org/10.1103/PhysRevLett.95.210503} {\bibfield  {journal}
  {\bibinfo  {journal} {Phys. Rev. Lett.}\ }\textbf {\bibinfo {volume} {95}},\
  \bibinfo {pages} {210503} (\bibinfo {year} {2005})}\BibitemShut {NoStop}%
\bibitem [{\citenamefont {Klimov}\ \emph {et~al.}(2018)\citenamefont {Klimov},
  \citenamefont {Kelly}, \citenamefont {Chen}, \citenamefont {Neeley},
  \citenamefont {Megrant}, \citenamefont {Burkett}, \citenamefont {Barends},
  \citenamefont {Arya}, \citenamefont {Chiaro}, \citenamefont {Chen},
  \citenamefont {Dunsworth}, \citenamefont {Fowler}, \citenamefont {Foxen},
  \citenamefont {Gidney}, \citenamefont {Giustina}, \citenamefont {Graff},
  \citenamefont {Huang}, \citenamefont {Jeffrey}, \citenamefont {Lucero},
  \citenamefont {Mutus}, \citenamefont {Naaman}, \citenamefont {Neill},
  \citenamefont {Quintana}, \citenamefont {Roushan}, \citenamefont {Sank},
  \citenamefont {Vainsencher}, \citenamefont {Wenner}, \citenamefont {White},
  \citenamefont {Boixo}, \citenamefont {Babbush}, \citenamefont {Smelyanskiy},
  \citenamefont {Neven},\ and\ \citenamefont {Martinis}}]{Klimov_2018}%
  \BibitemOpen
  \bibfield  {author} {\bibinfo {author} {\bibfnamefont {P.~V.}\ \bibnamefont
  {Klimov}}, \bibinfo {author} {\bibfnamefont {J.}~\bibnamefont {Kelly}},
  \bibinfo {author} {\bibfnamefont {Z.}~\bibnamefont {Chen}}, \bibinfo {author}
  {\bibfnamefont {M.}~\bibnamefont {Neeley}}, \bibinfo {author} {\bibfnamefont
  {A.}~\bibnamefont {Megrant}}, \bibinfo {author} {\bibfnamefont
  {B.}~\bibnamefont {Burkett}}, \bibinfo {author} {\bibfnamefont
  {R.}~\bibnamefont {Barends}}, \bibinfo {author} {\bibfnamefont
  {K.}~\bibnamefont {Arya}}, \bibinfo {author} {\bibfnamefont {B.}~\bibnamefont
  {Chiaro}}, \bibinfo {author} {\bibfnamefont {Y.}~\bibnamefont {Chen}},
  \bibinfo {author} {\bibfnamefont {A.}~\bibnamefont {Dunsworth}}, \bibinfo
  {author} {\bibfnamefont {A.}~\bibnamefont {Fowler}}, \bibinfo {author}
  {\bibfnamefont {B.}~\bibnamefont {Foxen}}, \bibinfo {author} {\bibfnamefont
  {C.}~\bibnamefont {Gidney}}, \bibinfo {author} {\bibfnamefont
  {M.}~\bibnamefont {Giustina}}, \bibinfo {author} {\bibfnamefont
  {R.}~\bibnamefont {Graff}}, \bibinfo {author} {\bibfnamefont
  {T.}~\bibnamefont {Huang}}, \bibinfo {author} {\bibfnamefont
  {E.}~\bibnamefont {Jeffrey}}, \bibinfo {author} {\bibfnamefont
  {E.}~\bibnamefont {Lucero}}, \bibinfo {author} {\bibfnamefont {J.~Y.}\
  \bibnamefont {Mutus}}, \bibinfo {author} {\bibfnamefont {O.}~\bibnamefont
  {Naaman}}, \bibinfo {author} {\bibfnamefont {C.}~\bibnamefont {Neill}},
  \bibinfo {author} {\bibfnamefont {C.}~\bibnamefont {Quintana}}, \bibinfo
  {author} {\bibfnamefont {P.}~\bibnamefont {Roushan}}, \bibinfo {author}
  {\bibfnamefont {D.}~\bibnamefont {Sank}}, \bibinfo {author} {\bibfnamefont
  {A.}~\bibnamefont {Vainsencher}}, \bibinfo {author} {\bibfnamefont
  {J.}~\bibnamefont {Wenner}}, \bibinfo {author} {\bibfnamefont {T.~C.}\
  \bibnamefont {White}}, \bibinfo {author} {\bibfnamefont {S.}~\bibnamefont
  {Boixo}}, \bibinfo {author} {\bibfnamefont {R.}~\bibnamefont {Babbush}},
  \bibinfo {author} {\bibfnamefont {V.~N.}\ \bibnamefont {Smelyanskiy}},
  \bibinfo {author} {\bibfnamefont {H.}~\bibnamefont {Neven}},\ and\ \bibinfo
  {author} {\bibfnamefont {J.~M.}\ \bibnamefont {Martinis}},\ }\bibfield
  {title} {\emph {\bibinfo {title} {Fluctuations of energy-relaxation times in
  superconducting qubits}},\ }\href
  {https://doi.org/10.1103/PhysRevLett.121.090502} {\bibfield  {journal}
  {\bibinfo  {journal} {Phys. Rev. Lett.}\ }\textbf {\bibinfo {volume} {121}},\
  \bibinfo {pages} {090502} (\bibinfo {year} {2018})}\BibitemShut {NoStop}%
\bibitem [{\citenamefont {Schl\"or}\ \emph {et~al.}(2019)\citenamefont
  {Schl\"or}, \citenamefont {Lisenfeld}, \citenamefont {M\"uller},
  \citenamefont {Bilmes}, \citenamefont {Schneider}, \citenamefont {Pappas},
  \citenamefont {Ustinov},\ and\ \citenamefont {Weides}}]{Schlor_2019}%
  \BibitemOpen
  \bibfield  {author} {\bibinfo {author} {\bibfnamefont {S.}~\bibnamefont
  {Schl\"or}}, \bibinfo {author} {\bibfnamefont {J.}~\bibnamefont {Lisenfeld}},
  \bibinfo {author} {\bibfnamefont {C.}~\bibnamefont {M\"uller}}, \bibinfo
  {author} {\bibfnamefont {A.}~\bibnamefont {Bilmes}}, \bibinfo {author}
  {\bibfnamefont {A.}~\bibnamefont {Schneider}}, \bibinfo {author}
  {\bibfnamefont {D.~P.}\ \bibnamefont {Pappas}}, \bibinfo {author}
  {\bibfnamefont {A.~V.}\ \bibnamefont {Ustinov}},\ and\ \bibinfo {author}
  {\bibfnamefont {M.}~\bibnamefont {Weides}},\ }\bibfield  {title} {\emph
  {\bibinfo {title} {Correlating decoherence in transmon qubits: Low frequency
  noise by single fluctuators}},\ }\href
  {https://doi.org/10.1103/PhysRevLett.123.190502} {\bibfield  {journal}
  {\bibinfo  {journal} {Phys. Rev. Lett.}\ }\textbf {\bibinfo {volume} {123}},\
  \bibinfo {pages} {190502} (\bibinfo {year} {2019})}\BibitemShut {NoStop}%
\bibitem [{\citenamefont {Burnett}\ \emph {et~al.}(2019)\citenamefont
  {Burnett}, \citenamefont {Bengtsson}, \citenamefont {Scigliuzzo},
  \citenamefont {Niepce}, \citenamefont {Kudra}, \citenamefont {Delsing},\ and\
  \citenamefont {Bylander}}]{Burnett_2019}%
  \BibitemOpen
  \bibfield  {author} {\bibinfo {author} {\bibfnamefont {J.~J.}\ \bibnamefont
  {Burnett}}, \bibinfo {author} {\bibfnamefont {A.}~\bibnamefont {Bengtsson}},
  \bibinfo {author} {\bibfnamefont {M.}~\bibnamefont {Scigliuzzo}}, \bibinfo
  {author} {\bibfnamefont {D.}~\bibnamefont {Niepce}}, \bibinfo {author}
  {\bibfnamefont {M.}~\bibnamefont {Kudra}}, \bibinfo {author} {\bibfnamefont
  {P.}~\bibnamefont {Delsing}},\ and\ \bibinfo {author} {\bibfnamefont
  {J.}~\bibnamefont {Bylander}},\ }\bibfield  {title} {\emph {\bibinfo {title}
  {Decoherence benchmarking of superconducting qubits}},\ }\href
  {http://dx.doi.org/10.1038/s41534-019-0168-5} {\bibfield  {journal} {\bibinfo
   {journal} {npj Quantum Information}\ }\textbf {\bibinfo {volume} {5}}
  (\bibinfo {year} {2019})}\BibitemShut {NoStop}%
\bibitem [{\citenamefont {Müller}\ \emph {et~al.}(2019)\citenamefont
  {Müller}, \citenamefont {Cole},\ and\ \citenamefont
  {Lisenfeld}}]{Muller_2019}%
  \BibitemOpen
  \bibfield  {author} {\bibinfo {author} {\bibfnamefont {C.}~\bibnamefont
  {Müller}}, \bibinfo {author} {\bibfnamefont {J.~H.}\ \bibnamefont {Cole}},\
  and\ \bibinfo {author} {\bibfnamefont {J.}~\bibnamefont {Lisenfeld}},\
  }\bibfield  {title} {\emph {\bibinfo {title} {Towards understanding
  two-level-systems in amorphous solids: insights from quantum circuits}},\
  }\href {https://doi.org/10.1088/1361-6633/ab3a7e} {\bibfield  {journal}
  {\bibinfo  {journal} {Reports on Progress in Physics}\ }\textbf {\bibinfo
  {volume} {82}},\ \bibinfo {pages} {124501} (\bibinfo {year}
  {2019})}\BibitemShut {NoStop}%
\bibitem [{\citenamefont {Matityahu}\ \emph {et~al.}(2016)\citenamefont
  {Matityahu}, \citenamefont {Shnirman}, \citenamefont {Schön},\ and\
  \citenamefont {Schechter}}]{Matityahu_2016}%
  \BibitemOpen
  \bibfield  {author} {\bibinfo {author} {\bibfnamefont {S.}~\bibnamefont
  {Matityahu}}, \bibinfo {author} {\bibfnamefont {A.}~\bibnamefont {Shnirman}},
  \bibinfo {author} {\bibfnamefont {G.}~\bibnamefont {Schön}},\ and\ \bibinfo
  {author} {\bibfnamefont {M.}~\bibnamefont {Schechter}},\ }\bibfield  {title}
  {\emph {\bibinfo {title} {Decoherence of a quantum two-level system by
  spectral diffusion}},\ }\href {https://doi.org/10.1103/PhysRevB.93.134208}
  {\bibfield  {journal} {\bibinfo  {journal} {Physical Review B}\ }\textbf
  {\bibinfo {volume} {93}},\ \bibinfo {pages} {134208} (\bibinfo {year}
  {2016})}\BibitemShut {NoStop}%
\bibitem [{\citenamefont {Schechter}\ \emph {et~al.}(2018)\citenamefont
  {Schechter}, \citenamefont {Nalbach},\ and\ \citenamefont
  {Burin}}]{Schechter_2018}%
  \BibitemOpen
  \bibfield  {author} {\bibinfo {author} {\bibfnamefont {M.}~\bibnamefont
  {Schechter}}, \bibinfo {author} {\bibfnamefont {P.}~\bibnamefont {Nalbach}},\
  and\ \bibinfo {author} {\bibfnamefont {A.~L.}\ \bibnamefont {Burin}},\
  }\bibfield  {title} {\emph {\bibinfo {title} {Nonuniversality and strongly
  interacting two-level systems in glasses at low temperatures}},\ }\href
  {https://doi.org/10.1088/1367-2630/aac930} {\bibfield  {journal} {\bibinfo
  {journal} {New Journal of Physics}\ }\textbf {\bibinfo {volume} {20}},\
  \bibinfo {pages} {063048} (\bibinfo {year} {2018})}\BibitemShut {NoStop}%
\bibitem [{\citenamefont {Yu}\ \emph {et~al.}(2022)\citenamefont {Yu},
  \citenamefont {Matityahu}, \citenamefont {Rosen}, \citenamefont {Hung},
  \citenamefont {Maksymov}, \citenamefont {Burin}, \citenamefont {Schechter},\
  and\ \citenamefont {Osborn}}]{Yu_2022}%
  \BibitemOpen
  \bibfield  {author} {\bibinfo {author} {\bibfnamefont {L.}~\bibnamefont
  {Yu}}, \bibinfo {author} {\bibfnamefont {S.}~\bibnamefont {Matityahu}},
  \bibinfo {author} {\bibfnamefont {Y.~J.}\ \bibnamefont {Rosen}}, \bibinfo
  {author} {\bibfnamefont {C.-C.}\ \bibnamefont {Hung}}, \bibinfo {author}
  {\bibfnamefont {A.}~\bibnamefont {Maksymov}}, \bibinfo {author}
  {\bibfnamefont {A.~L.}\ \bibnamefont {Burin}}, \bibinfo {author}
  {\bibfnamefont {M.}~\bibnamefont {Schechter}},\ and\ \bibinfo {author}
  {\bibfnamefont {K.~D.}\ \bibnamefont {Osborn}},\ }\bibfield  {title} {\emph
  {\bibinfo {title} {Experimentally revealing anomalously large dipoles in the
  dielectric of a quantum circuit}},\ }\href
  {https://doi.org/10.1038/s41598-022-21256-7} {\bibfield  {journal} {\bibinfo
  {journal} {Scientific Reports}\ }\textbf {\bibinfo {volume} {12}},\ \bibinfo
  {pages} {16960} (\bibinfo {year} {2022})}\BibitemShut {NoStop}%
\bibitem [{\citenamefont {Churkin}\ \emph {et~al.}(2014)\citenamefont
  {Churkin}, \citenamefont {Barash},\ and\ \citenamefont
  {Schechter}}]{Churkin_2014}%
  \BibitemOpen
  \bibfield  {author} {\bibinfo {author} {\bibfnamefont {A.}~\bibnamefont
  {Churkin}}, \bibinfo {author} {\bibfnamefont {D.}~\bibnamefont {Barash}},\
  and\ \bibinfo {author} {\bibfnamefont {M.}~\bibnamefont {Schechter}},\
  }\bibfield  {title} {\emph {\bibinfo {title} {Nonhomogeneity of the density
  of states of tunneling two-level systems at low energies}},\ }\href
  {https://doi.org/10.1103/PhysRevB.89.104202} {\bibfield  {journal} {\bibinfo
  {journal} {Physical Review B}\ }\textbf {\bibinfo {volume} {89}},\ \bibinfo
  {pages} {104202} (\bibinfo {year} {2014})}\BibitemShut {NoStop}%
\bibitem [{Y_R()}]{Y_Rosen_preparation}%
  \BibitemOpen
  \href@noop {} {}\bibinfo {note} {Y. Rosen et. al., in
  preparation}\BibitemShut {NoStop}%
\bibitem [{\citenamefont {Shalibo}\ \emph {et~al.}(2010)\citenamefont
  {Shalibo}, \citenamefont {Rofe}, \citenamefont {Shwa}, \citenamefont
  {Zeides}, \citenamefont {Neeley}, \citenamefont {Martinis},\ and\
  \citenamefont {Katz}}]{PhysRevLett.105.177001}%
  \BibitemOpen
  \bibfield  {author} {\bibinfo {author} {\bibfnamefont {Y.}~\bibnamefont
  {Shalibo}}, \bibinfo {author} {\bibfnamefont {Y.}~\bibnamefont {Rofe}},
  \bibinfo {author} {\bibfnamefont {D.}~\bibnamefont {Shwa}}, \bibinfo {author}
  {\bibfnamefont {F.}~\bibnamefont {Zeides}}, \bibinfo {author} {\bibfnamefont
  {M.}~\bibnamefont {Neeley}}, \bibinfo {author} {\bibfnamefont {J.~M.}\
  \bibnamefont {Martinis}},\ and\ \bibinfo {author} {\bibfnamefont
  {N.}~\bibnamefont {Katz}},\ }\bibfield  {title} {\emph {\bibinfo {title}
  {Lifetime and coherence of two-level defects in a josephson junction}},\
  }\href {https://doi.org/10.1103/PhysRevLett.105.177001} {\bibfield  {journal}
  {\bibinfo  {journal} {Phys. Rev. Lett.}\ }\textbf {\bibinfo {volume} {105}},\
  \bibinfo {pages} {177001} (\bibinfo {year} {2010})}\BibitemShut {NoStop}%
\bibitem [{Thermal cycling was rigorously analyzed in Ref.
  \cite{PhysRevLett.105.177001}, yet the re-initiation of the TLS excitation
  spectrum at $\approx 20$K but not at $\approx 1$K is typical to quantum
  circuits()}]{note_2024}%
  \BibitemOpen
  Thermal cycling was rigorously analyzed in Ref.
  \cite{PhysRevLett.105.177001}, yet the re-initiation of the TLS excitation
  spectrum at $\approx 20$K but not at $\approx 1$K is typical to quantum
  circuits,\ \href@noop {} {}\BibitemShut {NoStop}%
\bibitem [{\citenamefont {Churkin}\ \emph {et~al.}(2023)\citenamefont
  {Churkin}, \citenamefont {Gabdank}, \citenamefont {Burin},\ and\
  \citenamefont {Schechter}}]{churkin2023strain}%
  \BibitemOpen
  \bibfield  {author} {\bibinfo {author} {\bibfnamefont {A.}~\bibnamefont
  {Churkin}}, \bibinfo {author} {\bibfnamefont {I.}~\bibnamefont {Gabdank}},
  \bibinfo {author} {\bibfnamefont {A.~L.}\ \bibnamefont {Burin}},\ and\
  \bibinfo {author} {\bibfnamefont {M.}~\bibnamefont {Schechter}},\ }\bibfield
  {title} {\emph {\bibinfo {title} {The strain gap in a system of weakly and
  strongly interacting two-level systems}},\ }\href
  {https://doi.org/10.1140/epjs/s11734-023-00982-7} {\bibfield  {journal}
  {\bibinfo  {journal} {Eur. Phys. J. Spec. Top.}\ }\textbf {\bibinfo {volume}
  {232}},\ \bibinfo {pages} {3483} (\bibinfo {year} {2023})}\BibitemShut
  {NoStop}%
\bibitem [{Sup()}]{SupplementalMaterial}%
  \BibitemOpen
  \href@noop {} {}\bibinfo {note} {See Supplemental Material at [URL will be
  inserted by publisher] for details.}\BibitemShut {Stop}%
\bibitem [{\citenamefont {Watson}(1995)}]{Watson_1995}%
  \BibitemOpen
  \bibfield  {author} {\bibinfo {author} {\bibfnamefont {S.~K.}\ \bibnamefont
  {Watson}},\ }\bibfield  {title} {\emph {\bibinfo {title} {Tunneling states in
  crystals with large random strains}},\ }\href
  {https://doi.org/10.1103/PhysRevLett.75.1965} {\bibfield  {journal} {\bibinfo
   {journal} {Phys. Rev. Lett.}\ }\textbf {\bibinfo {volume} {75}},\ \bibinfo
  {pages} {1965--1968} (\bibinfo {year} {1995})}\BibitemShut {NoStop}%
\bibitem [{\citenamefont {Liu}\ \emph {et~al.}(1998)\citenamefont {Liu},
  \citenamefont {Vu}, \citenamefont {Pohl}, \citenamefont {Schiettekatte},\
  and\ \citenamefont {Roorda}}]{Liu_1998}%
  \BibitemOpen
  \bibfield  {author} {\bibinfo {author} {\bibfnamefont {X.}~\bibnamefont
  {Liu}}, \bibinfo {author} {\bibfnamefont {P.~D.}\ \bibnamefont {Vu}},
  \bibinfo {author} {\bibfnamefont {R.~O.}\ \bibnamefont {Pohl}}, \bibinfo
  {author} {\bibfnamefont {F.}~\bibnamefont {Schiettekatte}},\ and\ \bibinfo
  {author} {\bibfnamefont {S.}~\bibnamefont {Roorda}},\ }\bibfield  {title}
  {\emph {\bibinfo {title} {Generation of low-energy excitations in silicon}},\
  }\href {https://doi.org/10.1103/PhysRevLett.81.3171} {\bibfield  {journal}
  {\bibinfo  {journal} {Phys. Rev. Lett.}\ }\textbf {\bibinfo {volume} {81}},\
  \bibinfo {pages} {3171--3174} (\bibinfo {year} {1998})}\BibitemShut {NoStop}%
\bibitem [{\citenamefont {Topp}\ \emph {et~al.}(1999)\citenamefont {Topp},
  \citenamefont {Thompson},\ and\ \citenamefont {Pohl}}]{Topp_1999}%
  \BibitemOpen
  \bibfield  {author} {\bibinfo {author} {\bibfnamefont {K.~A.}\ \bibnamefont
  {Topp}}, \bibinfo {author} {\bibfnamefont {E.}~\bibnamefont {Thompson}},\
  and\ \bibinfo {author} {\bibfnamefont {R.~O.}\ \bibnamefont {Pohl}},\
  }\bibfield  {title} {\emph {\bibinfo {title} {Glasslike excitations in
  chemically disordered crystals: Alkali-earth lanthanum fluoride mixed
  crystals}},\ }\href {https://doi.org/10.1103/PhysRevB.60.898} {\bibfield
  {journal} {\bibinfo  {journal} {Phys. Rev. B}\ }\textbf {\bibinfo {volume}
  {60}},\ \bibinfo {pages} {898--908} (\bibinfo {year} {1999})}\BibitemShut
  {NoStop}%
\bibitem [{\citenamefont {Topp}\ and\ \citenamefont {Pohl}(2002)}]{Topp_2002}%
  \BibitemOpen
  \bibfield  {author} {\bibinfo {author} {\bibfnamefont {K.~A.}\ \bibnamefont
  {Topp}}\ and\ \bibinfo {author} {\bibfnamefont {R.~O.}\ \bibnamefont
  {Pohl}},\ }\bibfield  {title} {\emph {\bibinfo {title} {Tunneling states in
  strained alkali-halide crystals containing ${\mathrm{cn}}^{\ensuremath{-}}$
  ions}},\ }\href {https://doi.org/10.1103/PhysRevB.66.064204} {\bibfield
  {journal} {\bibinfo  {journal} {Phys. Rev. B}\ }\textbf {\bibinfo {volume}
  {66}},\ \bibinfo {pages} {064204} (\bibinfo {year} {2002})}\BibitemShut
  {NoStop}%
\bibitem [{\citenamefont {Pohl}\ \emph {et~al.}(1999)\citenamefont {Pohl},
  \citenamefont {Liu;},\ and\ \citenamefont {Crandall}}]{POHL_opinion_1999}%
  \BibitemOpen
  \bibfield  {author} {\bibinfo {author} {\bibfnamefont {R.}~\bibnamefont
  {Pohl}}, \bibinfo {author} {\bibfnamefont {X.}~\bibnamefont {Liu;}},\ and\
  \bibinfo {author} {\bibfnamefont {R.}~\bibnamefont {Crandall}},\ }\bibfield
  {title} {\emph {\bibinfo {title} {Lattice vibrations of disordered solids}},\
  }\href {https://doi.org/https://doi.org/10.1016/S1359-0286(99)00028-5}
  {\bibfield  {journal} {\bibinfo  {journal} {Current Opinion in Solid State
  and Materials Science}\ }\textbf {\bibinfo {volume} {4}},\ \bibinfo {pages}
  {281--287} (\bibinfo {year} {1999})}\BibitemShut {NoStop}%
\bibitem [{\citenamefont {Treacy}\ and\ \citenamefont
  {Borisenko}(2012)}]{Treacy_2012}%
  \BibitemOpen
  \bibfield  {author} {\bibinfo {author} {\bibfnamefont {M.~M.~J.}\
  \bibnamefont {Treacy}}\ and\ \bibinfo {author} {\bibfnamefont {K.~B.}\
  \bibnamefont {Borisenko}},\ }\bibfield  {title} {\emph {\bibinfo {title} {The
  local structure of amorphous silicon}},\ }\href
  {https://doi.org/10.1126/science.1214780} {\bibfield  {journal} {\bibinfo
  {journal} {Science}\ }\textbf {\bibinfo {volume} {335}},\ \bibinfo {pages}
  {950--953} (\bibinfo {year} {2012})}\BibitemShut {NoStop}%
\bibitem [{V_I()}]{V_Iaia_preparation}%
  \BibitemOpen
  \href@noop {} {}\bibinfo {note} {V. Iaia et. al., in preparation}\BibitemShut
  {NoStop}%
\bibitem [{\citenamefont {Rieger}\ \emph {et~al.}(2023)\citenamefont {Rieger},
  \citenamefont {Gunzler}, \citenamefont {Spiecker}, \citenamefont {Paluch},
  \citenamefont {Winkel}, \citenamefont {Hahn}, \citenamefont {Hohmann},
  \citenamefont {Bacher}, \citenamefont {Wernsdorfer},\ and\ \citenamefont
  {Pop}}]{Rieger2023}%
  \BibitemOpen
  \bibfield  {author} {\bibinfo {author} {\bibfnamefont {D.}~\bibnamefont
  {Rieger}}, \bibinfo {author} {\bibfnamefont {S.}~\bibnamefont {Gunzler}},
  \bibinfo {author} {\bibfnamefont {M.}~\bibnamefont {Spiecker}}, \bibinfo
  {author} {\bibfnamefont {P.}~\bibnamefont {Paluch}}, \bibinfo {author}
  {\bibfnamefont {P.}~\bibnamefont {Winkel}}, \bibinfo {author} {\bibfnamefont
  {L.}~\bibnamefont {Hahn}}, \bibinfo {author} {\bibfnamefont {J.~K.}\
  \bibnamefont {Hohmann}}, \bibinfo {author} {\bibfnamefont {A.}~\bibnamefont
  {Bacher}}, \bibinfo {author} {\bibfnamefont {W.}~\bibnamefont
  {Wernsdorfer}},\ and\ \bibinfo {author} {\bibfnamefont {I.~M.}\ \bibnamefont
  {Pop}},\ }\bibfield  {title} {\emph {\bibinfo {title} {Granular aluminium
  nanojunction fluxonium qubit}},\ }\href
  {https://doi.org/https://doi.org/10.1038/s41563-022-01417-9} {\bibfield
  {journal} {\bibinfo  {journal} {Nature Materials}\ }\textbf {\bibinfo
  {volume} {22}},\ \bibinfo {pages} {194} (\bibinfo {year} {2023})}\BibitemShut
  {NoStop}%
\bibitem [{\citenamefont {de~Graaf}\ \emph {et~al.}(2020)\citenamefont
  {de~Graaf}, \citenamefont {Faoro}, \citenamefont {Ioffe}, \citenamefont
  {Mahashabde}, \citenamefont {Burnett}, \citenamefont {Lindström},
  \citenamefont {Kubatkin}, \citenamefont {Danilov},\ and\ \citenamefont
  {Tzalenchuk}}]{deGraaf2020}%
  \BibitemOpen
  \bibfield  {author} {\bibinfo {author} {\bibfnamefont {S.~E.}\ \bibnamefont
  {de~Graaf}}, \bibinfo {author} {\bibfnamefont {L.}~\bibnamefont {Faoro}},
  \bibinfo {author} {\bibfnamefont {L.~B.}\ \bibnamefont {Ioffe}}, \bibinfo
  {author} {\bibfnamefont {S.}~\bibnamefont {Mahashabde}}, \bibinfo {author}
  {\bibfnamefont {J.~J.}\ \bibnamefont {Burnett}}, \bibinfo {author}
  {\bibfnamefont {T.}~\bibnamefont {Lindström}}, \bibinfo {author}
  {\bibfnamefont {S.~E.}\ \bibnamefont {Kubatkin}}, \bibinfo {author}
  {\bibfnamefont {A.~V.}\ \bibnamefont {Danilov}},\ and\ \bibinfo {author}
  {\bibfnamefont {A.~Y.}\ \bibnamefont {Tzalenchuk}},\ }\bibfield  {title}
  {\emph {\bibinfo {title} {Two-level systems in superconducting quantum
  devices due to trapped quasiparticles}},\ }\href
  {https://doi.org/10.1126/sciadv.abc5055} {\bibfield  {journal} {\bibinfo
  {journal} {Science Advances}\ }\textbf {\bibinfo {volume} {6}},\ \bibinfo
  {pages} {eabc5055} (\bibinfo {year} {2020})}\BibitemShut {NoStop}%
\bibitem [{\citenamefont {Gr\"unhaupt}\ \emph {et~al.}(2018)\citenamefont
  {Gr\"unhaupt}, \citenamefont {Maleeva}, \citenamefont {Skacel}, \citenamefont
  {Calvo}, \citenamefont {Levy-Bertrand}, \citenamefont {Ustinov},
  \citenamefont {Rotzinger}, \citenamefont {Monfardini}, \citenamefont
  {Catelani},\ and\ \citenamefont {Pop}}]{Grunhaupt2018}%
  \BibitemOpen
  \bibfield  {author} {\bibinfo {author} {\bibfnamefont {L.}~\bibnamefont
  {Gr\"unhaupt}}, \bibinfo {author} {\bibfnamefont {N.}~\bibnamefont
  {Maleeva}}, \bibinfo {author} {\bibfnamefont {S.~T.}\ \bibnamefont {Skacel}},
  \bibinfo {author} {\bibfnamefont {M.}~\bibnamefont {Calvo}}, \bibinfo
  {author} {\bibfnamefont {F.}~\bibnamefont {Levy-Bertrand}}, \bibinfo {author}
  {\bibfnamefont {A.~V.}\ \bibnamefont {Ustinov}}, \bibinfo {author}
  {\bibfnamefont {H.}~\bibnamefont {Rotzinger}}, \bibinfo {author}
  {\bibfnamefont {A.}~\bibnamefont {Monfardini}}, \bibinfo {author}
  {\bibfnamefont {G.}~\bibnamefont {Catelani}},\ and\ \bibinfo {author}
  {\bibfnamefont {I.~M.}\ \bibnamefont {Pop}},\ }\bibfield  {title} {\emph
  {\bibinfo {title} {Loss mechanisms and quasiparticle dynamics in
  superconducting microwave resonators made of thin-film granular aluminum}},\
  }\href {https://doi.org/10.1103/PhysRevLett.121.117001} {\bibfield  {journal}
  {\bibinfo  {journal} {Phys. Rev. Lett.}\ }\textbf {\bibinfo {volume} {121}},\
  \bibinfo {pages} {117001} (\bibinfo {year} {2018})}\BibitemShut {NoStop}%
\bibitem [{\citenamefont {Vepsäläinen}\ \emph {et~al.}(2020)\citenamefont
  {Vepsäläinen}, \citenamefont {Karamlou}, \citenamefont {Orrell},
  \citenamefont {Dogra}, \citenamefont {Loer}, \citenamefont {Vasconcelos},
  \citenamefont {Kim}, \citenamefont {Melville}, \citenamefont {Niedzierlski},
  \citenamefont {Yoder}, \citenamefont {Gustavsson}, \citenamefont {Formaggio},
  \citenamefont {VanDevender},\ and\ \citenamefont {Oliver}}]{Vepsalainen2020}%
  \BibitemOpen
  \bibfield  {author} {\bibinfo {author} {\bibfnamefont {A.}~\bibnamefont
  {Vepsäläinen}}, \bibinfo {author} {\bibfnamefont {A.}~\bibnamefont
  {Karamlou}}, \bibinfo {author} {\bibfnamefont {J.}~\bibnamefont {Orrell}},
  \bibinfo {author} {\bibfnamefont {A.}~\bibnamefont {Dogra}}, \bibinfo
  {author} {\bibfnamefont {B.}~\bibnamefont {Loer}}, \bibinfo {author}
  {\bibfnamefont {F.}~\bibnamefont {Vasconcelos}}, \bibinfo {author}
  {\bibfnamefont {D.}~\bibnamefont {Kim}}, \bibinfo {author} {\bibfnamefont
  {A.}~\bibnamefont {Melville}}, \bibinfo {author} {\bibfnamefont
  {B.}~\bibnamefont {Niedzierlski}}, \bibinfo {author} {\bibfnamefont
  {J.}~\bibnamefont {Yoder}}, \bibinfo {author} {\bibfnamefont
  {S.}~\bibnamefont {Gustavsson}}, \bibinfo {author} {\bibfnamefont
  {J.}~\bibnamefont {Formaggio}}, \bibinfo {author} {\bibfnamefont
  {B.}~\bibnamefont {VanDevender}},\ and\ \bibinfo {author} {\bibfnamefont
  {W.}~\bibnamefont {Oliver}},\ }\bibfield  {title} {\emph {\bibinfo {title}
  {Impact of ionizing radiation on superconducting qubit coherence}},\ }\href
  {https://doi.org/10.1038/s41586-020-2619-8} {\bibfield  {journal} {\bibinfo
  {journal} {Nature}\ }\textbf {\bibinfo {volume} {584}},\ \bibinfo {pages}
  {551} (\bibinfo {year} {2020})}\BibitemShut {NoStop}%
\bibitem [{\citenamefont {Wilen}\ \emph {et~al.}(2021)\citenamefont {Wilen},
  \citenamefont {Abdullah}, \citenamefont {Kurinsky}, \citenamefont {Stanford},
  \citenamefont {Cardani}, \citenamefont {D’Imperio}, \citenamefont {Tomei},
  \citenamefont {Faoro}, \citenamefont {Ioffe}, \citenamefont {Liu},
  \citenamefont {Opremcak}, \citenamefont {Christensen}, \citenamefont
  {DuBois}, ,\ and\ \citenamefont {McDermott}}]{Wilen2021}%
  \BibitemOpen
  \bibfield  {author} {\bibinfo {author} {\bibfnamefont {C.~D.}\ \bibnamefont
  {Wilen}}, \bibinfo {author} {\bibfnamefont {S.}~\bibnamefont {Abdullah}},
  \bibinfo {author} {\bibfnamefont {N.~A.}\ \bibnamefont {Kurinsky}}, \bibinfo
  {author} {\bibfnamefont {C.}~\bibnamefont {Stanford}}, \bibinfo {author}
  {\bibfnamefont {L.}~\bibnamefont {Cardani}}, \bibinfo {author} {\bibfnamefont
  {G.}~\bibnamefont {D’Imperio}}, \bibinfo {author} {\bibfnamefont
  {C.}~\bibnamefont {Tomei}}, \bibinfo {author} {\bibfnamefont
  {L.}~\bibnamefont {Faoro}}, \bibinfo {author} {\bibfnamefont {L.~B.}\
  \bibnamefont {Ioffe}}, \bibinfo {author} {\bibfnamefont {C.~H.}\ \bibnamefont
  {Liu}}, \bibinfo {author} {\bibfnamefont {A.}~\bibnamefont {Opremcak}},
  \bibinfo {author} {\bibfnamefont {B.~G.}\ \bibnamefont {Christensen}},
  \bibinfo {author} {\bibfnamefont {J.~L.}\ \bibnamefont {DuBois}}, ,\ and\
  \bibinfo {author} {\bibfnamefont {R.}~\bibnamefont {McDermott}},\ }\bibfield
  {title} {\emph {\bibinfo {title} {Correlated charge noise and relaxation
  errors in superconducting qubits}},\ }\href
  {https://doi.org/10.1038/s41586-021-03557-5} {\bibfield  {journal} {\bibinfo
  {journal} {Nature}\ }\textbf {\bibinfo {volume} {594}},\ \bibinfo {pages}
  {369} (\bibinfo {year} {2021})}\BibitemShut {NoStop}%
\bibitem [{\citenamefont {Thorbeck}\ \emph {et~al.}(2023)\citenamefont
  {Thorbeck}, \citenamefont {Eddins}, \citenamefont {Lauer}, \citenamefont
  {McClure},\ and\ \citenamefont {Carroll}}]{Thorbeck2023}%
  \BibitemOpen
  \bibfield  {author} {\bibinfo {author} {\bibfnamefont {T.}~\bibnamefont
  {Thorbeck}}, \bibinfo {author} {\bibfnamefont {A.}~\bibnamefont {Eddins}},
  \bibinfo {author} {\bibfnamefont {I.}~\bibnamefont {Lauer}}, \bibinfo
  {author} {\bibfnamefont {D.~T.}\ \bibnamefont {McClure}},\ and\ \bibinfo
  {author} {\bibfnamefont {M.}~\bibnamefont {Carroll}},\ }\bibfield  {title}
  {\emph {\bibinfo {title} {Two-level-system dynamics in a superconducting
  qubit due to background ionizing radiation}},\ }\href
  {https://doi.org/10.1103/PRXQuantum.4.020356} {\bibfield  {journal} {\bibinfo
   {journal} {PRX Quantum}\ }\textbf {\bibinfo {volume} {4}},\ \bibinfo {pages}
  {020356} (\bibinfo {year} {2023})}\BibitemShut {NoStop}%
\end{thebibliography}%

\onecolumngrid

\begin{center}
    \rule{0.8\linewidth}{1pt}
\end{center}
\vspace{0.5cm}

\section*{Supplementary Materials}

\makeatother
\renewcommand{\thesubsection}{\Roman{subsection}}

\renewcommand{\theequation}{S\arabic{equation}}

\renewcommand{\thefigure}{S\arabic{figure}}

\setcounter{equation}{0}
\setcounter{figure}{0}
\preprint{APS/123-QED}

\subsection{\label{sec:TwoTLS} The Two-TLS model in disordered lattices and in amorphous solids}

The low-temperature universal phenomena in amorphous solids is in fact a broad phenomena in strongly disordered systems, observed also in disordered crystals, polymers, and quasicrystals \cite{pohl2002low}. For example, KBr$_{1-x}$:CN$_x$, perhaps the archetypal disordered lattice exhibiting the universal phenomena, has, for $0.25<x<0.7$, and below $1$K, specific heat $\propto T^{\alpha}$ with $\alpha \sim 1$, thermal conductivity $\propto T^{\beta}$ with $\beta \sim 2$, and nearly temperature independent internal friction. The observation of the low-temperature universal phenomena in a large variety of disordered lattices led to the idea that the phenomena are not related to the lack of long range order in amorphous solids, but rather to the strong local disorder, shared by both amorphous solids and disordered lattices. Additional evidence for this notion was given by experiment on ion-implanted crystalline silicon. Implantation by Boron ions at high dosage leads to amorphicity, while implantation by Silicon ions keeps long-range lattice order intact. Yet, both samples show at high implantation dose the same universal phenomena, suggesting that local disorder rather than the absence of long-range order dictate the low-temperature phenomena in strongly disordered and amorphous solids. This led the authors to argue that the defects in the crystal should be used to model the excitations in the amorphous silicon, rather than the amorphous structure itself \cite{Liu_1998}.
Disordered lattices are advantageous in the study of the low-temperature universal phenomena for various reasons. TLSs in disordered solids can be identified, e.g. CN dipoles in KBr$_{1-x}$:CN$_x$, and disordered lattices can be carefully manipulated to address physical questions. A prominent example is the debate of whether the universal phenomena is dictated by interactions between the tunneling defects, or by the strain disorder. To address this question a series of experiments studied highly strained disordered lattice, e.g. KBr$_{0.5}$:Cl$_{0.5}$, with a small concentration of tunneling defects, e.g. CN impurities \cite{Watson_1995,Topp_1999,Topp_2002}, where it was shown that universal phenomena, linear in the concentration of the tunneling defects, appear, despite the smallness of interactions between the dilute defects, pointing to the strong strain disorder as dictating the universal low-temperature phenomena. Such experiments led Pohl {\it et.al.,} \cite{POHL_opinion_1999} to argue that the nature of the defects, and the cause for their saturation density, should be explored in disordered crystals, rather than in amorphous solids.

Following this proposal, the Two-TLS model was suggested in an attempt to describe properly the low-energy properties of disordered lattices in a model which will then allow making predictions to be tested also in amorphous solids, to support its general validity in describing the low-temperature properties in strongly disordered and amorphous solids. Considering for example KBr$_{1-x}$:CN$_x$, natural candidates for tunneling states are the CN defects, which, however, can tunnel either by $90^{\circ}$ rotation or by $180^{\circ}$ flip. As a result of inversion symmetry, CN rotations and flips, however, have a very different strain interaction. In particular a single CN impurity in an otherwise pure lattice has, to first order, zero interaction of the flip TLS with the strain. In a strongly disordered KBr$_{1-x}$:CN$_x$ system, with $0.25<x<0.7$, inversion symmetry is not purely maintained. Yet, the interactions of CN flips with the strain are proportional to the deviations from local inversion symmetry, which are, at near neighbor distance, of the order of $3\%$ in strongly disordred lattices, as well as in amorphous solids \cite{Treacy_2012}. As a result, CN flips and rotations constitute two sets of tunneling TLSs, differing by their interaction with the strain, denoted $\gamma_w$ and $\gamma_s$ respectively, with $g\equiv \gamma_w/\gamma_s \approx 0.03$. Thus, the Two-TLS model generalizes the interaction of TLSs with the strain given by the STM

\begin{equation}
H_{\rm TLS-ph} \;=\; \sum_j \sum_{\alpha,\beta} \;
\gamma^{\alpha \beta} \tau_j^z \; \frac{\partial u_j^{\alpha}}{\partial
{\bf r}_j^{\beta}} \, .
\label{impurityphonon}
\end{equation}
to include the two sets of TLSs differing by their strain interaction, reading:

\begin{equation}
H^{\rm tot}_{\rm TLS-ph} \;=\; \sum_{j,\alpha,\beta} \;
\gamma_{\rm s}^{\alpha \beta} S_j^z \; \frac{\partial u_j^{\alpha}}{\partial
{\bf r}_j^{\beta}} + \sum_{j',\alpha,\beta} \;
\gamma_{\rm w}^{\alpha \beta} \tau_{j'}^z \; \frac{\partial u_{j'}^{\alpha}}{\partial
{\bf r}_{j'}^{\beta}} .
\label{Stauphonon}
\end{equation}

Here $u_j^{\alpha}$ denotes the $\alpha=x,y,z$ component of the phononic amplitude at ${\bf r}_j$. This bi-modality of the strain interaction was confirmed numerically using ab-initio and DFT calculations in \cite{Gaita_2011}. For the STM form of the TLS-phonon interaction, Eq.~(\ref{impurityphonon}), integration out of the phonons results in an effective Hamiltonian for TLS-TLS interactions, given by

\begin{align}
\mathcal{H}_{\tau int} = -\sum_{i\neq j}\frac{1}{2} J_{ij}^{\tau \tau}\tau_{i}\tau_{j} \, .
\label{Eq:TLSInteractions}
\end{align}
and including the static disorder induced random field one obtains Eq.~(3) in the main part.
For the generalized Two-TLS interaction model, Eq.~(\ref{Stauphonon}), similar integration out of the phonons results in the effective TLS-TLS interactions given in Eq.~(1) in the main part, and repeated here for convenience:

\begin{align}
\mathcal{H}_{s\tau} = -\sum_{i\neq j}[\frac{1}{2} J_{ij}^{ss}S_{i}S_{j} + J_{ij}^{s\tau}S_{i}\tau_{j} + \frac{1}{2} J_{ij}^{\tau \tau}\tau_{i}\tau_{j}].
\label{Eq:TwoTLSSupp}
\end{align}

The form of the above interaction Hamiltonian has essential consequences on the low energy DOS of the $\tau$-TLSs and of the $S$-TLSs. For the $\tau$-TLSs, the dominate energy is given by its interaction with the $S$-TLSs, which are mostly frozen at low energies, leading to a similar effective Hamiltonian to that of the STM [Eq. (3) in the main part]. However, the $S$-TLSs are pseudo-gapped at low energies, consequence of their interactions with other $S$-TLSs, and with the more abundant at low energies $\tau$-TLSs, resulting in power-low energy dependent $S$-TLS DOS at low energies \cite{churkin2023strain}. This form of the DOS was calculated analytically for the model Hamiltonian in Eq.(\ref{Eq:TwoTLSSupp}) \cite{churkin2023strain}, as well as from first principles for the KBr$_{1-x}$:CN$_x$ system, using hybrid molecular statics and Monte-Carlo numerical technique \cite{Churkin_2014}, with excellent agreement between the results for the real system and for the model.

As a result of the scarceness of the $S$-TLSs at low energies, their effect on most equilibrium properties is negligible. The $\tau$-TLSs are equivalent to the standard TLSs in the STM. Yet, the smallness of their interaction with the strain dictates the effective disorder exerted on the $\tau$-TLSs to be $g J_0 \approx 10$K, i.e. about $g$ times smaller than assumed by the STM. Even more importantly, their DOS at low energies is proportional to $(\gamma_w \gamma_s)^{-1}$, leading to smallness of $\gamma_w/\gamma_s$ of the tunneling strength, which together with the standard logarithmic factor of the ratio of the maximal to minimal TLS tunneling amplitudes $\ln(\Delta_0^{max}/\Delta_0^{min})$, produces the smallness and universality of the tunneling strength. Thus, within the Two-TLS model, the universality of the low-temperature properties of strongly disordered and amorphous solids lies in the rather universal value of the typical strain in such systems, which dictates the value of $g \approx 0.03$.

Unlike the case in disordered lattices, the identification of the tunneling states in amorphous solids is a formidable task. Thus, to address the validity of the Two-TLS model for amorphous solids one resorts to making predictions unique to the Two-TLS model, and checking them against experimental results. For the Two-TLS model to give predictions which are different from that of the STM one needs to have properties with a dominant contribution of the $S$-TLSs, despite their scarceness at low energies in equilibrium. Examples are the fast switching of thermal $S$-TLSs dictating echo dephasing of resonant TLSs \cite{Matityahu_2016}, loss of resonators under fast bias sweeping which brings $S$-TLSs from high energies into resonance \cite{Yu_2022,Y_Rosen_preparation} and thermal cycling experiments, as discussed in the current manuscript.

\subsection{\label{sec:level2}RATIONALE AND JUSTIFICATIONS FOR THE EXTRAPOLATION METHODOLOGY}

In the main part we conduct extrapolation on the system size to assess the overall impact of the heating and cooling cycles on the excitation energies of the $\tau$-TLSs. Within the extrapolation process we utilize the probability of flip for the $S$-TLSs, $P_s(n)$, obtained from simulations performed on a small-scale system. In this procedure we assume that the sizes used for the direct simulation, $L_i=10,12,16$, suffice to produce relevant distributions $P_s(n)$ for extrapolation.
This approximation is supported by the results presented in Fig. 1 (b) in the main part, yet we would like to look into this approximation in more detail. While TLS-TLS interaction is long range, its value at distances larger than $L/2=6$ is roughly $5$K. Since $S$-TLSs have excitation energies that are typically much larger than $5$K, and are pseudo-gapped at low energies ${\lesssim}10$K
we expect that upon thermal cycling to temperature ${\lesssim}20$K the correlation of $S$-TLS flips of distances larger than $L/2=6$ to be very small.

\begin{figure*}[h]

\centering
\includegraphics[width=1\textwidth]{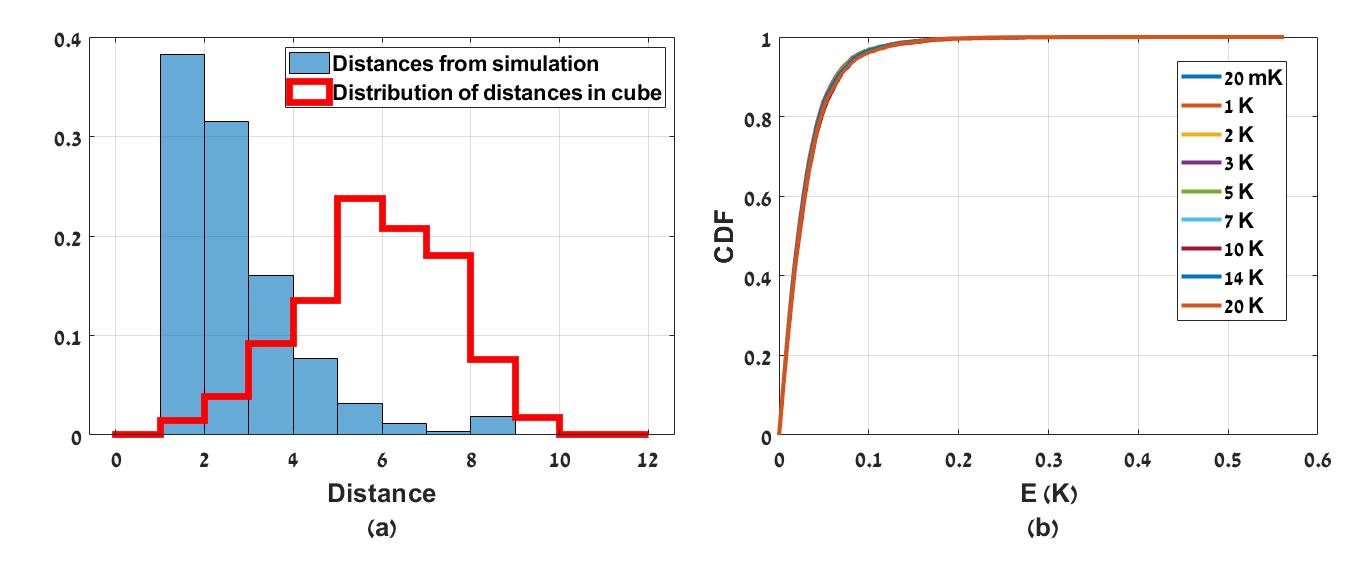}
\caption{(a) Comparison between the distance distribution of TLS pairs in a $12^{3}$ cube with a particle density of $\rho = 1/2$, and the distribution of the distances between the S-TLSs pairs which flipped, in the case where only 2 S-TLSs flipped. (b) Cumulative distribution function (CDF) of the initial energies of $\tau$-TLSs that flipped after thermal cycling to different heating temperatures, in simulations where none of the S-TLSs flipped.}
\label{Fig:SPairs}
\end{figure*}

Support to this assertion is given in Fig. \ref{Fig:SPairs}(a), where we plot for $L_i=12$ the distribution of distances between flipped $S$-TLS pairs in those samples where exactly two $S$-TLSs are flipped following thermal cycling. Results clearly indicate that the effect of correlated flips of $S$-TLSs is a local phenomenon.

Another point to consider is that in the extrapolation process, we solely consider the number of $S$-TLSs that flipped in each cube due to the heating and cooling cycle, while disregarding the contribution of the flipped $\tau$-TLSs to the change of the excitation energies of other $\tau$-TLSs. The justification for this neglect can be divided into two scenarios. First, in cubes where at least a single $S$-TLS has flipped, the relative contribution to the change of $\tau$-TLS excitation energies by flipped $\tau$-TLSs is negligible, as $J_{0}^{\tau\tau} \ll J_{0}^{S\tau}$.
Then, in cubes where no $S$-TLSs have flipped as a result of the heating and cooling cycle, flipped $\tau$-TLSs are predominantly thermal TLSs, causing small jitter in the excitation energies of other $\tau$-TLSs, yet no long-time change in the mean of their excitation energies. In Fig. \ref{Fig:SPairs}(b) we plot the integrated number of flipped $\tau$-TLSs as function of their excitation energies, indeed confirming that they are predominantly thermal.

\subsection{\label{sec:level2} CONVERGENCE OF THE EXTRAPOLATION RESULTS}

\begin{figure}[h]

\includegraphics[width=0.5\textwidth]{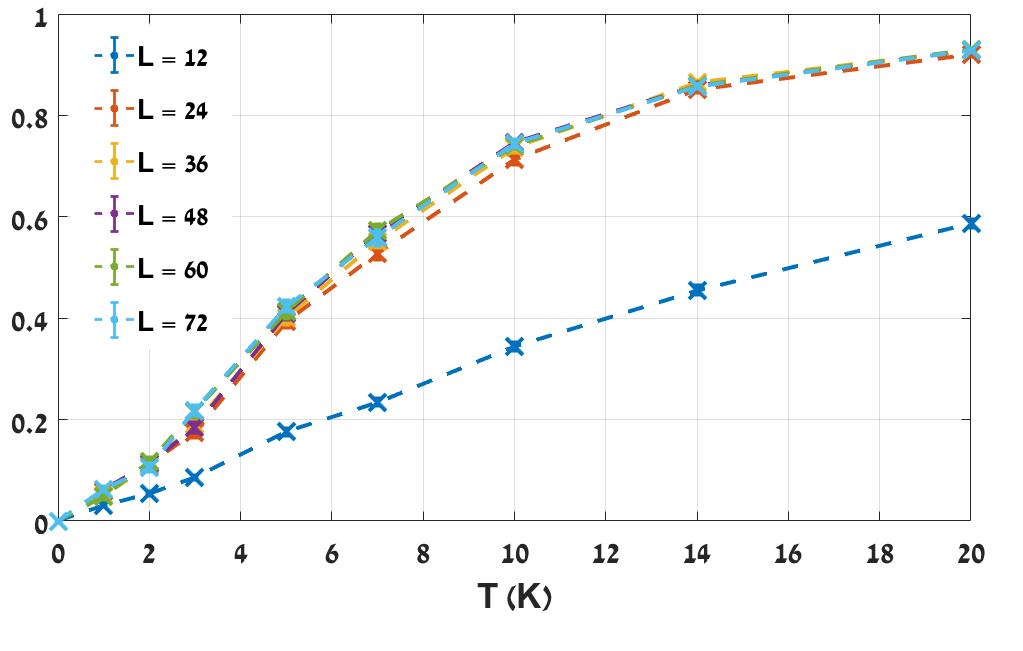}
\caption{Fraction of $\tau$-TLSs whose excitation energy has changed by more than $\Delta E$ = 0.1K. $L=12$ from direct calculation. Results for larger sizes are based on the probability distribution $P_{s}$ generated from the simulation for $L=12$.}
\label{Fig:extrapolation_12_to_60_de_01}
\end{figure}

In the main part we present the key results regarding the number of $\tau$-TLSs that experienced a change in their excitation energy beyond a specific value referred to as $\Delta E$. We also provide a detailed explanation of the extrapolation process. Fig \ref{Fig:extrapolation_12_to_60_de_01} illustrates the results of the extrapolation for increasing sample sizes until convergence is reached.

\begin{figure}[h]

\includegraphics[width=0.5\textwidth]{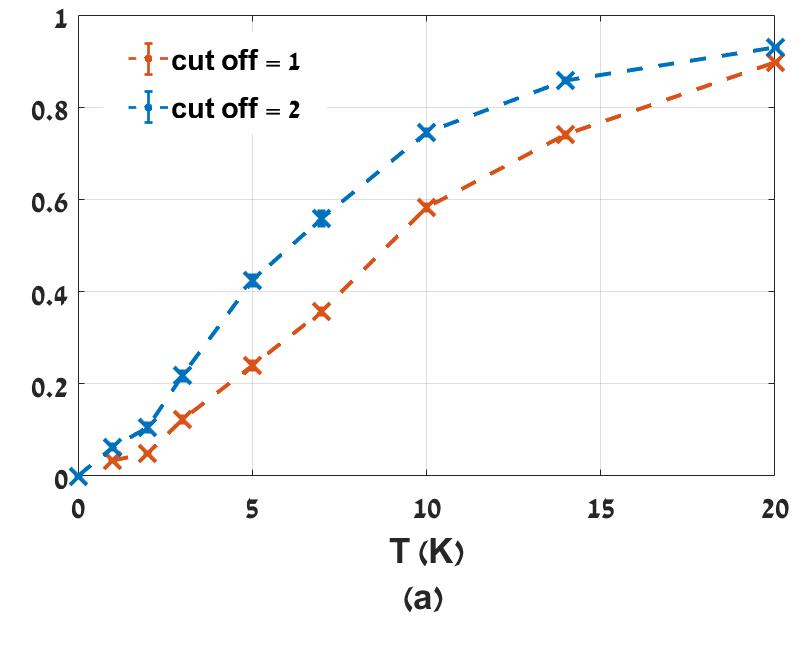}
\includegraphics[width=0.5\textwidth]{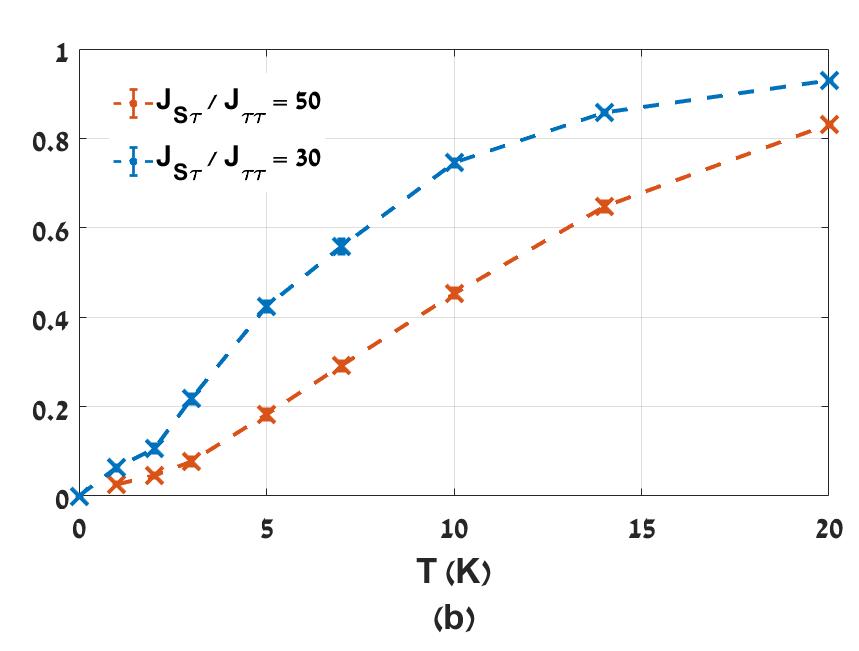}
\includegraphics[width=0.5\textwidth]{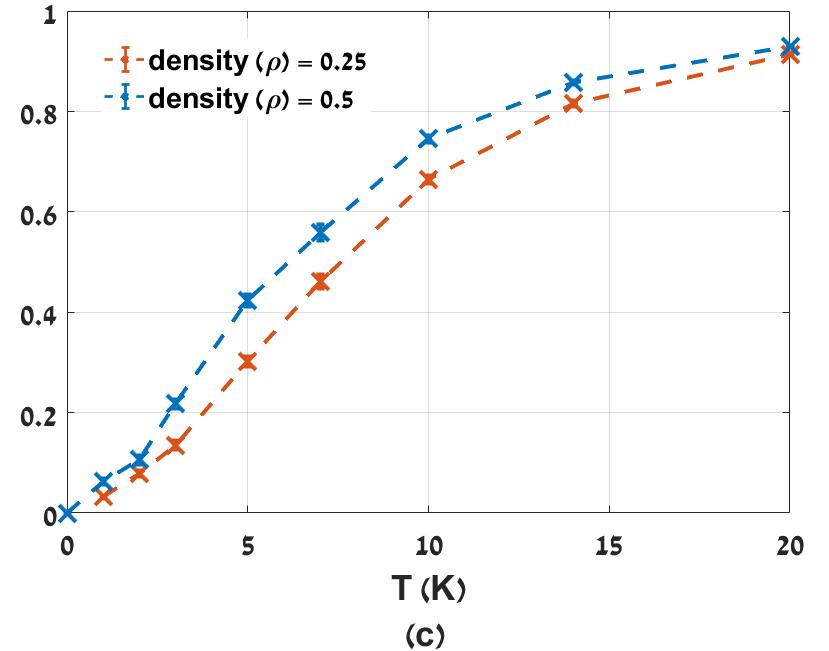}
\caption{Fraction of $\tau$-TLSs whose excitation energies change by more than $\Delta E = 0.1 \text{K}$ for (a) different cutoff values, (b) different strain interaction ratios, and (c) different $\rho$ (TLS densities). Presented results are after extrapolation to large sizes until conversion (as in Fig. 1(b) in the main part), using $P_{s}$ deduced from simulations of size $L = 12$ ($L = 15$ for $\rho = 0.25$, to maintain the same number of particles).}
\label{Fig:parameters}
\end{figure}

\subsection{\label{sec:level2} SENSITIVITY TO PARAMETERS CHANGE}
Our simulations for the Two-TLS model involve several parameters: $\tilde{a}$, the interaction cutoff distance, $g$, the ratio between the strain coupling strengths of $\tau$-TLSs and that of $S$-TLSs, and the density $\rho$ at which we embed the TLSs in the lattice. Here, we compare the simulation results for variations of these parameters while keeping the $\tau$-TLS DOS unvaried. The latter is Gaussian \cite{churkin2023strain}, with width given by

\begin{align}
    E_{\text{typ}}^{\tau} = \sqrt{\frac{16 \pi}{3} \frac{\rho}{\tilde{a}^{3}}} \, gJ_0.
    \label{eq:E_type}
\end{align}
which we keep constant by adjusting the parameter \( J_0 \).

Results are presented in Fig. \ref{Fig:parameters} following extrapolation to large system sizes until convergence. The calculations use \( P_s \) generated from simulations of \( L = 12 \) lattice sizes (\( L = 15 \) for \( \rho = 0.25 \), to maintain the same number of particles in the sample). The main conclusion drawn from our results here is that for parameter changes which are within the regime compatible with the Two-TLS model our qualitative results remain intact: at cycling temperatures of $\approx 20$ K, the excitation energies of almost all \(\tau\)-TLSs change by more than 0.1 K, whereas at cycling temperatures of $\approx 1$K, only a small number of \(\tau\)-TLSs exhibit such a change. Yet, the quantitative changes as function of the relevant parameters may turn out useful in extracting the model parameters from experiments.

Let us now discuss the quantitative dependence on simulation parameters of the number of \(\tau\)-TLSs that upon thermal cycling change their excitation energy by more than $0.1$K.
While we keep the $\tau$-TLS DOS unchanged, the change of parameters has two main effects on the system: (i) changing the interaction strength between the TLSs (ii) changing the DOS of the $S$-TLSs.
With regard to the former, the change of $S - \tau$ interaction strength changes the radius of influence of a flipped $S$ TLS, given by \cite{churkin2023strain}

\begin{align}
    R &= \sqrt[3]{\frac{4 J_0 g}{\Delta E} + \tilde{a}^3} \, .
    \label{eq:R}
\end{align}
For \( \Delta E \ll 4J_0 g \), we obtain
\begin{align}
    N &\approx \frac{16 \pi J_0 g}{3 \Delta E} \rho \, .
    \label{eq:N_approx}
\end{align}
For a specific \( \Delta E \), this simplifies to
\begin{align}
    N &\propto J_0 g \rho
    \label{eq:N_propto}
\end{align}
and substituting \( J_0 \) from equation \ref{eq:E_type}, we get
\begin{align}
    N &\propto \rho^{1/2} \tilde{a}^{3/2} \, .
    \label{eq:N_substituted}
\end{align}
As a result, the number of \(\tau\)-TLSs that change their excitation energy at a given \( \Delta E \) given a single flipped $S$-TLS increases with enhanced
\( \tilde{a} \) and with enhanced \( \rho \). A competing effect is the reduction of the low energy DOS of $S$-TLSs with enhanced \( \tilde{a} \) and \( \rho \). Yet, this effect is subdominant. For example, the change of $\tilde{a}$ from $2$ to $1$ results in an enhancement of the $S$-TLS DOS by only $\approx 50\%$ (see Fig. $4$ in Ref. \cite{churkin2023strain}). Consequently, the overall dependence of the number of $\tau$-TLSs which experience a given change in their excitation energy upon thermal cycling is monotonically increasing as function of \( \tilde{a} \) and as function of \( \rho \). The effect becomes less dominant with the increase of the cycling temperature, as regions within the radius of influence of distinct flipped $S$-TLSs begin to overlap.

Changing the value of $g$, however, while maintaining \( E_{\text{typ}}^{\tau} \) at a constant value by keeping $J_0 \propto 1/g$ results in a significant effect on the DOS of the $S$-TLSs. The reduction of $g$ and the corresponding increase of $J_0$ increases the typical value of $S$-TLS energies, and appreciably decreases the DOS of the $S$-TLSs at low energies. At the same time, from equation \eqref{eq:N_substituted} we find that the number of \(\tau\)-TLSs affected by a single $S$-TLS flip does not depend on \( g \). We thus find a decrease in the total number of affected \(\tau\)-TLSs due to a reduction in \( g \).

\[
\]
[1] A. Churkin, I. Gabdank, A. L. Burin, and M. Schechter, Eur. Phys. J. Spec. Top. {\bf 232}, 3483 (2023).

\end{document}